\documentclass[12pt,halfline,a4paper]{ouparticle}

\usepackage{natbib}
\usepackage{booktabs}
\usepackage{threeparttable}
\usepackage{amsmath}
\usepackage{amssymb}
\usepackage{longtable}
\usepackage{graphicx}
\usepackage{caption}
\usepackage{subcaption}
\usepackage{tabularx}
\usepackage{listings}
\usepackage{titlesec}
\usepackage[titletoc,title]{appendix}
\renewcommand{\thetable}{\Roman{table}}
\renewcommand{\thefigure}{\Roman{figure}}
\renewcommand{\thesection}{\Roman{section}}
\renewcommand{\thesubsection}{\thesection.\Alph{subsection}}
\begin{document}

\title{GPT takes the SAT: Tracing changes in\break {Test Difficulty and Students' Math Performance}}

\author{%
    \name{Vikram K Suresh}
    \address{University of Cincinnati, Economics; Computer Science}
    \email{krishnvv@mail.uc.edu}
    \and
    \name{Saannidhya Rawat}
    \address{University of Cincinnati, Economics, Cincinnati, Ohio}
    \email{rawatsa@mail.uc.edu}}

\abstract{Scholastic Aptitude Test (SAT) is crucial for college admissions but its effectiveness and relevance are increasingly questioned. This paper enhances Synthetic Control methods by introducing \textit{Transformed Control}, a novel method that employs Large Language Models (LLMs) powered by Artificial Intelligence to generate control groups. We utilize OpenAI's API to generate a control group where GPT-4, or ChatGPT, takes multiple SATs annually from 2008 to 2023. This control group helps analyze shifts in SAT math difficulty over time, starting from the baseline year of 2008. Using parallel trends, we calculate the Average Difference in Scores (ADS) to assess changes in high school students' math performance. Our results indicate a significant decrease in the difficulty of the SAT math section over time, alongside a decline in students' math performance. The analysis shows a 71-point drop in the rigor of SAT math from 2008 to 2023, with student performance decreasing by 36 points, resulting in a 107-point total divergence in average student math performance. We investigate possible mechanisms for this decline in math proficiency, such as changing university selection criteria, increased screen time, grade inflation, and worsening adolescent mental health. Disparities among demographic groups show a 104-point drop for White students, 84 points for Black students, and 53 points for Asian students. Male students saw a 117-point reduction, while female students had a 100-point decrease.
}




\maketitle


\pagebreak

\section{Introduction} \label{sec:intro}

The Scholastic Aptitude Test (SAT) has long been a cornerstone of college admissions in the United States, serving as a standardized measure of academic readiness. Nevertheless, recent literature has re-assessed its efficacy and fairness in evaluating student potential and post-college success \citep{Chetty2023-mf}.

This study employs a novel approach to investigate the changing complexity of the Scholastic Aptitude Test (SAT) Math section over time and its potential impact on high school students' readiness for the rigors of higher education. The urgency and significance of this research stem from the increasing scrutiny of standardized tests like the SAT, as universities reconsider their role in the admissions process amidst concerns about equity, access, and the comprehensive assessment of student capabilities. Our study not only responds to these critical educational policy questions but also pioneers the application of Artificial Intelligence (AI) in evaluating and interpreting changes in standardized testing over time. By leveraging OpenAI's Large Language Model (LLM) powered AI solutions, we introduce a novel 'Transformed Control' methodology, setting a new precedent in the synthetic control literature and offering a fresh lens through which to assess educational standards and student outcomes. Our methodology is grounded in the use of LLM agents as a stable benchmark against which to measure the SAT Math section's difficulty across years. This approach allows us to bypass traditional experimental constraints and ethical concerns associated with human subjects, presenting a scalable and unbiased means to examine the test's evolution from 2008 to 2023. The findings of our investigation are both revealing and concerning, pointing to a noticeable decline in the rigor of SAT Math questions over the examined period. This trend is paralleled by a significant drop in students' mathematical proficiency, as inferred from their performance relative to the LLM agent's scores. Such results underscore not only a shift in the nature of the SAT but also raise questions about the broader educational ecosystem's ability to nurture and elevate the mathematical preparedness among high school students.

The College Board which conducts the Scholastic Aptitude Test (SAT) announced in 2017 that preparation guides can boost the scores, it may have signalled an arms race with the test takers \citep{CollegeBoard2017}. For the SAT to continue being an effective sorting tool for the universities, it must remain in step with test takers who have access to numerous preparation guides. On the other hand, 80\% of 4-year degree granting universities have made the SAT an optional submission criterion for applicants \citep{FairTest}. 85 universities did not consider an SAT score for fall 2023 admissions even if it was submitted. \cite{Bennett2021} suggests that while test-blind admissions minimally affect application volume, they could enhance access for the socioeconomically disadvantaged, although their impact on long-term post-graduation outcomes is uncertain. \cite{Saboe2019} find test optional policies have no effect on racial and socioeconomic diversity or gender ratio and minimal effect on application volume.

\cite{Chetty2023-mf} offers fresh perspective on the admissions process and student outcomes from 12 elite universities, the Ivy-League, Duke, MIT, Stanford and University of Chicago. The applicant volume from the top 1\% income households is disproportionately higher compared to the middle- and lower-income households given similar SAT scores. The factors include legacy admissions, athletic scholarships and non-academic credentials from elite private high schools, recommendations and extra-curricular activities. Importantly, the SAT is found to be a better predictor of long-term outcomes post graduation than non-academic credentials. Chetty’s analysis concludes, by over-weighting non-academic credentials and preference for high-income legacy applicants, the elite universities have created a bottleneck for the American middle and low-income classes from reaching the upper echelons of leadership positions in society.
 
To determine if universities can still rely on the SAT to admit students, we test the hypothesis, are the current SAT exams more rigorous than before? If so, the SAT is not only keeping up with the preparation guides available to test takers but still providing an important sorting tool to the universities. The emergence of AI solutions through instruction-tuned LLMs give us an opportunity to compare the rigorous of the SAT exam over time.

Traditionally, to conduct such an experiment, the research design must include the same students or identical sample of students each taking the SAT exam from different years. To control for the idiosyncratic variations, the students must take the exam in identical hermetically sealed conditions. The resulting distribution of scores given each exam can help us evaluate the possible difference in the SAT exam from different periods. The increasing complexity of the experiment makes it an impractical proposition. AI solutions provide models that operate as an independent arbiter of the exams, not constrained by the physiological and psychological limitations of human test takers in experimental conditions. Further, they alleviate the ethical and moral concerns in dealing with human subjects. In this way, AI augments our analysis by allowing us to construct a separate control group that does not have these aforementioned shortcomings associated with a traditional experiment.

Although the question of significance of SAT in college admissions to foster equitable access to higher education has been studied by other researchers in recent years (see \cite{bleemer2021affirmative}, \cite{mountjoy2020returns}), this paper offers a unique perspective by using the state-of-the-art AI technology to facilitate the debate of relevance of standardized tests in the current college admissions process. The paper focuses on several strands of literature in this paper and engages with the literature on three main fronts.

Firstly, we address the literature surrounding the role of standardized tests in college admissions. The previous work related to standardized tests has been focused on heterogeneous impacts of standardized tests on different sections of the population (see \cite{reardon2018widening}, \cite{card2007racial}, \cite{hoekstra2009effect}). None of the papers in the literature quantify the possible changes in the difficulty of SAT exam. The exam is assumed to be of fixed difficulty level only requiring score re-centering as recommended by the College Board to account for format changes and syllabus updates when inter-temporal comparisons are performed. As the SAT scores do not expire, minor adjustments are deemed sufficient when comparing the SAT scores over time. Our results show that SAT has become less rigorous over time.

Secondly, we contribute to the literature surrounding novel and heterodox approaches in Economics profession. As far as we know, this is the first paper that uses Large Language Models (LLMs) as a benchmark counterfactual. We use this counterfactual to compare against the math performance of high school students. We develop a control group consisting of GPT-4's average SAT Math score over time compared to the baseline year of 2008, which we call 'Transformed Control'. This control group acts in ways similar to a standard synthetic control group introduced by \cite{abadie2003economic}. However, instead of using different weights to form a comparable control group as done in the standard synthetic control literature, we use LLM powered AI to form our control group. The transformed control will then be used to assess evolving exam difficulty and math performance of high school students.

Lastly, we combine the application of a standard empirical estimation methods such as multiple period difference-in-difference method and Bayesian comparison of means to compute the estimates for the parameters of interest. We accommodate multi-period difference-in-difference method to our setting by introducing a measure that we refer to as Average Difference in Scores (ADS) and we use it to estimate the change in mathematical performance of high school students after controlling for change in exam difficulty over time.

The subsequent sections of this paper are organized in the following manner. Section \ref{sec:data} introduces the context of the SAT examination, highlighting significant modifications to its format that are recognized and addressed within this study. Additionally, this section elaborates on the SAT exam data utilized for evaluating GPT model scores. Section \ref{sec:EmpStrategy} delves into the empirical methodology applied for administering the SAT exams to Large Language Models (LLMs), alongside comparisons drawn between the two cohorts. Section \ref{sec:results} presents the findings of this research. Section \ref{sec:het} discusses heterogeneity analysis conducted, and Section \ref{sec:mech} offers some possible mechanisms. Section \ref{sec:conclusion} concludes.

\section{Background \& Data}  \label{sec:data}



\subsection{SAT Data}
To construct a robust counterfactual framework for evaluating LLM agents on SAT exams, we required a comprehensive dataset of SAT exams. We carefully curated these SAT exams from a variety of online sources specializing in SAT preparation, which provided access to an extensive collection of past SAT examinations and practice tests. We gathered these exams in PDF format and stored them in a secure digital repository, encompassing a chronological collection from 2008 to 2023.

After collecting the PDF files, we transcribed each SAT exam PDF into a structured comma-separated values (CSV) file. This CSV file included a comprehensive range of information pertinent to each exam question, such as the question text, options for multiple-choice questions (MCQs), the original source document for the question, section number, question number, type of question (MCQ or answer-type), and calculator usage policy. MCQs present a set of possible answers, requiring the examinee to select the most appropriate option. Answer type allowed a one-line input from the examinee. Given high performing multi-modal AI models that accept visual prompts were not available at the time of running this experiment, we excluded questions incorporating graphical or large tabular data from our analysis. This exclusion ensured compatibility with the text-only input capabilities of the GPT models under investigation.

In 2016, the SAT exam underwent significant changes. The format shifted from a total score of 2400 to 1600, aligning with the pre-2005 SAT format. The Math section retained its value of 800 points. Importantly, the revision eliminated the penalty for incorrect answers, encouraging students to attempt all questions without fear of point deduction for incorrect guesses. Additionally, the Math section saw a reduction in the number of sections from three to two and an increase in questions from 54 to 58. Before 2016, calculators were allowed for all math sections. Starting 2016, the Math section included both a calculator-permitted and a \textit{no calculator} section. It is important to acknowledge these format changes and how they might impact our study.

\vskip 0.3in

1. \textit{Score conversion tables}. In assessing performance on the SAT, one can evaluate either the \textit{raw} score, denoting the number of correctly answered questions, or the \textit{scaled} score, a transformed value ranging between 200 and 800, as standardized by the College Board. For the purpose of our analysis, we designate the raw score as the count of correct responses and the scaled score as the numerical value within the 200-800 range post conversion. To facilitate this transformation from raw to scaled scores, we utilized distinct conversion tables for the \textit{pre} and \textit{post} format change. In this paper, we refer to \textit{pre} period as all the years before and including 2016, while \textit{post} period as all the years starting 2017. Although these tables exhibit slight annual variations, our review of potential variances due to the use of a fixed table per period revealed minimal impact on our results. The tables used for score conversion from \textit{raw} to \textit{scaled} score is provided in the appendix.

Furthermore, in the context of SAT score analysis, it is essential to acknowledge that an identical scaled score from the \textit{pre} format change period, may not equate to the same level of performance in the \textit{post} format change period. For example, an SAT score of 700 in \textit{pre} period may not be equivalent to SAT score of 700 in the \textit{post} period. This discrepancy arises from differences in the underlying question count and the conversion methodologies employed for translating raw scores into the standardized 200-800 scaled score range. To address this, we utilized the concordance tables provided by the College Board, which facilitate an accurate mapping of scaled scores from the \textit{pre} period to those in the \textit{post} period. This mapping is instrumental in ensuring a valid and equivalent comparison of performance across the two periods, effectively neutralizing any variances attributable solely to the score conversion process. The conversion according to the prescribed concordance table rounds the score to the nearest multiple of 10. We instead use a simple linear model given the concordance table to map the \textit{pre} format change scores to the \textit{post} format change scores to account for intermediate values. We also provide the concordance table used in this study in the appendix.

\vskip 0.1in

2. \textit{Negative marking}. As previously stated, our study spans two eras, 2008-2016 known as \textit{pre} and 2017-2023 known as \textit{post} period. During \textit{pre} period, the exams implemented negative marking, unlike the 2017-2023 period. In analyzing the GPT models' performance, we opted not to apply negative scoring for incorrect responses when converting from raw score to scaled score. This decision might initially suggest an inflated assessment of GPT performance in the \textit{pre} period, as measured by scaled SAT scores. However, as our forthcoming sections demonstrate, incorporating negative marking for this period would likely have amplified the performance discrepancy between the two periods. Therefore, our analysis likely presents a conservative estimate, potentially downplaying the actual performance gap.

\vskip 0.1in

3. \textit{Syllabus change}. Although SAT Math section did not experience a major change in structure and syllabus as compared to the verbal section, we must emphasize certain differences in exam content between the two periods. Before 2016, the key focus areas were arithmetic, numbers and operations, algebra, functions, geometry and data analysis. Starting 2016, SAT exam expanded the range of topics further and included questions related to trigonometry and complex numbers. Also, there was a greater focus on data analysis, graphs and word problems and less emphasis on geometry-related questions. Furthermore, starting 2016, students started received sub-scores for sections labeled as \textit{Heart of Algebra}, \textit{Passport to Advanced Math} and \textit{Problem Solving and Data Analysis}. There were no such subsections before 2016. Just by looking at the syllabus, one could argue that the Math section became harder than before as it covered the number of topics covered increased, along with their difficulty level.

\subsection{Average student SAT scores data}
The College Board provides yearly reports with the population level SAT performance by the cohort of high school students taking the SAT exam. The data from these reports include average scores, total test takers and standard deviation in the scores for the mathematics test and the language and writing test. These reports are readily available for years 2016 onward through the College Board website. For the years prior to 2016, we collected the then released reports from the Internet Archive. We then compiled the necessary information about average performance of average high school student in the United States over time. The summary estimate of the average SAT performance and standard errors are provided in figure. We made the appropriate conversion using the concordance table and mapped the average scores from the \textit{pre} period to the \textit{post} period.

The concordance table provided by the College board is in multiples of 10. This requires rounding the average SAT score for exams in the \textit{pre} period to the nearest multiple of 10 before it can be mapped to the average SAT score based on the concordance table. Due to this, the direct conversion of SAT scores using the concordance table has certain limitations. For example, if an average SAT score is 514 in the year 2009 and the score is 515 in the year 2010, 514 would be rounded to 510 and 515 would be rounded to 520, before the concordance table can be used. To avoid this problem, we used a simple linear regression model to regress average SAT scores before and after the conversion in the concordance table and then linearly interpolate the average SAT scores, which we call the Concordance SAT scores.

The figure below shows the average SAT score of students by year, before and after the conversion, as well as after using the linear interpolation technique.


The red line represents the original average SAT scores for each year from 2008 to 2016. The green line represents the converted score obtained after rounding the scores in the red line to the nearest multiple of 10, and then applying the conversion in the concordance table. The green line represents the concordance using the linear interpolation technique explained above. The average SAT scores for each year from 2017 to 2023 remain unaltered and are directly available through the College Board reports.

After using the concordance table, we can see that the mean SAT score decline in the \textit{post} period compared to the \textit{pre} period. This may suggest that students are doing worse in the \textit{post} period exams than the \textit{pre} period exams. However, we are making the assumption that the underlying exam did not change in difficulty.

As mentioned earlier, for our analysis encompassing U.S., we used the average SAT score data provided by College Board for each year. A conspicuous limitation of using average SAT score for each year is the lack of number of observations available for the study, since we have one observation each for period from 2008 to 2016, and the same for period from 2017 to 2023. Granted this data limitation, we provide a number of supporting results to check if the treatment effect estimates are robust. Instead of relying on average SAT scores at U.S level, we use unweighted mean SAT score data for each state to estimate the parameters of interest. The state means thus provide us a viable national sample to compare against the LLM agent. Lastly, we pick one specific state out of the 50 states to perform school district-level analysis and estimate the parameters of interest for the state of Massachusetts. In all the cases, we find similar results.

\subsection{State-level SAT scores data}

For state-level SAT reports from 2016 onwards, we directly accessed the state-level reports available on the College Board's website. These reports provide comprehensive insights into the SAT performance of students on a state-by-state basis, enabling a detailed examination of trends and patterns in SAT scores across the United States. However, the availability of state-level SAT data prior to 2016 posed a unique challenge. To procure this historical data for the \textit{pre} period (2008-2016), we used the Internet Archive. This digital archive, renowned for its extensive collection of web pages archived over time, proved instrumental in retrieving past state-level SAT reports not readily accessible on the College Board's current website. Utilizing this, we systematically sourced and compiled state-level SAT score reports for each year from 2008 to 2016.

This approach of combining current and archived data sources ensures a comprehensive and continuous dataset spanning the entire duration of our study period. We used this data to construct national sample of unweighted SAT averages from the state-level data for the years considered.

\subsection{District-level SAT scores data}
We used publicly available school district level SAT data for the state of Massachusetts from the Massachusetts Department of Education. The data are available for all the school districts in the state for each year from 2004. The number of test takers, the average reading, writing and mathematics scores are provided by school district for a given academic year. We identified 228 unique school districts commonly present over all the years considered in the study. The distribution of the average scores for these school districts is provided in the figures \ref{fig:avg_stud_scores_mass} and \ref{fig:avg_stud_scores_mass_conv}. The distribution of average scores after conversion maps the \textit{pre} period scores to match the \textit{post} format change period.



\section{Empirical Strategy} \label{sec:EmpStrategy}
\subsection{Reduced form approach}

For this study, we have two parameters of interest. First, we want to estimate the change in difficulty of SAT math section. This tells us whether the exam has changed over the years or not in terms of its mathematical rigor. Second, we want to estimate the change in mathematical aptitude of high school students as measured by Math SAT score. We explain our empirical strategy to estimate these parameters below.


1. \textit{Transformed Control}. In order to estimate the change in difficulty of the SAT math section relative to the baseline period, we extend the principles of Synthetic Control by \cite{abadie2003economic} and introduce the paradigm of ‘Transformed Control’.  Large Language Models (LLMs) like OpenAI’s ChatGPT have revolutionized Natural Language Processing (NLP) tasks. These models are built using the ‘Transformer Neural Network’ architecture and refined with Reinforcement Learning through Human Feedback (RLHF). The models are trained using unsupervised learning on vast corpus of text data. The text data are broken down into chunks called tokens, which are processed in parallel leveraging the Transformer architecture to understand the context of the word with reference to all other words in the sentence. The LLMs thus act as predictive engines for tokenized output which we can perceive as human language. The model performance is evaluated on metrics relevant to the specific task such as text-completion, question-answering, code-completion, etc. We do not explicate the LLMs further, \cite{korinek2023generative} provides a comprehensive guide on their day-to-day use case for economists. As these models are trained on increasing amounts of human generated text data and are continually improved, they become more adept at understanding and responding to queries. This progress opens new and sophisticated applications for these models. We highlight one such application within social sciences: generating data for what we refer to as ‘Transformed Control’ in our research design. The LLMs have been used for direct causal reasoning by \cite{kıcıman2023causal}, given a causal question to understand the models’ capacity to accurately identify casual factors. Our approach is distinct from this procedure as we are using the LLM to generate control group data to enrich the inference about the treatment group. Although the control in this design does not serve for causal inference yet, it paves the way for incorporating modern tools into social science research.

The 'transformed control' has similar advantages that are prominent with synthetic control methods which are highlighted by \cite{Cunningham2021CausalInference}. Our method precludes extrapolation, the comparison with high school student SAT outcomes is based on the LLM agent's performance in SAT exams from the same year. The counterfactual constructed relies on questions students themselves faced in that year. Secondly, given the collection of SAT questions from exams in a particular year, the LLM agent performs independently of student outcomes in the same year as it is a black-box neutrally answering the questions. The black box does not learn from previous attempts as each API call to the model is independent. The black box uses the knowledge in its training data to generate an appropriate answer to the question provided. This ensures the LLM agent's responses are not influenced by 'peeking' at the student outcomes as each evaluation of the SAT exam is independent and unbiased. While standardized tests have been used to identify trends in student performance over time, it is assumed the underlying exam is uniformly challenging. Independently evaluating the standardized tests would require large logistical undertaking. An identical control group must take the standardized tests to identify the bias introduced by the relative difficulty of the test taken by cohorts over the years. The LLM agent can instead act as the control and thus bridge the gap between qualitative and quantitative research. In our experiment the LLM by OpenAI, GPT-4 Turbo (GPT-4 here on) is the control unit. The algorithm provided in the Figure \ref{fig:gptflowchart_A} outlines the process of providing and receiving a response from the LLM agent.

Certain limitations should be considered before deploying LLMs as the control cohort in an experiment. Since different LLMs have different training data structure and size, the benchmarked performance varies among them. Model internal weights are often proprietary, the intellectual property is guarded by model providers and can vary arbitrarily based on their reinforcement training. While the safety protocols surrounding AI will remain contentious, the tool is too invaluable to be ignored by social sciences. The choice of LLM will therefore influence the transformed control data, we chose GPT-4 as it is the highest benchmarked model available through API at the time of writing \citep{openai-gpt4}. Additionally, the prompting technique affects the generated outcome as shown by \cite{gpt-brazil2023}. The model can significantly improve its performance when allowed to solve the questions using sequential reasoning. We constrained this capacity of the model in the experiment using a zero-shot prompting, limited token output and zero-temperature (see appendix \ref{sec:appxa} for details). The change in the model’s performance relative to the baseline year, gives the relative difficulty of the SAT. We assume this relative difficulty is LLM invariant as it is an evaluation of the underlying SAT exam over time by the same LLM agent and all agent specific effects (through their weights, training data size, prompt etc.) cancel out.

2. \textit{Performance Comparison}. Our second parameter of interest is the change in mathematical aptitude of the high school students. To estimate this parameter, we borrow estimation procedure from the multiple-period difference-in-difference framework as expounded in \cite{CallawaySantAnna2021}. Next, we explain how we borrow ideas and notation from \cite{CallawaySantAnna2021} to fit our empirical setting.

The LLM agent, GPT-4 has a fixed aptitude given the prompting technique is held constant. This constantly held agent takes multiple SAT exams from each year between 2008 and 2023. If the high school students similarly had unchanging mathematical aptitude, the expected change in performance of the students in the SAT exam is the estimated change in the underlying exam difficulty. We use the potential outcome notation for the high school students' SAT score measurements, $Y_{i,t}(0)$ is potential $i^{th}$ measurement in period $t$ for the unchanging high school students. While $Y_{i,t}(g)$ is the actual outcome measured for students. We formally define the $j^{th}$ exam score for the LLM agent taken in period $t$ as $T_{j,t}$. This allows us to construct a strong parallel trends like assumption as the following,
\begin{equation*}
    \mathbb{E}[Y_{t}(0) - Y_{t-1}(0)] = \mathbb{E}[T_{t} - T_{t-1}],\text{ where } 2008 < t \le 2023.
\end{equation*}

Since our 'transformed control' group is the unobserved parallel trend, the estimation mimics treatment effect estimation in a classical difference-in-difference approach without making causal claims. Additionally, under this parallel trends like assumption, the expected change in the LLM agent's score is the estimated change in difficulty of the SAT exam between the measurement periods. We borrow some notation from the estimation procedure described by \cite{CallawaySantAnna2021} and adopt for our use case. Our control provides a richer perspective about the difference in student performance, we refer to this estimand as the Average Difference in Scores (ADS).
\begin{equation*}
    ADS(\hat{t},t) = \mathbb{E}[Y_{t}(g)-Y_{t}(0)\text{ }|\text{ }\hat{t} \ ]
\end{equation*}
In the above equation, $Y_{t}(0)$ is the unobserved potential outcome. However, when we impose strong parallel trends, we can identify the ADS with reference to baseline year $\hat{t}=2008$,
\begin{align*}
    ADS(\hat{t},t) & =  \mathbb{E}[Y_{t}(g)-Y_{t}(0)\text{ }|\text{ }\hat{t} \ ] + \mathbb{E}(Y_{\hat{t}} (0) | \hat{t}) - \mathbb{E}(Y_{\hat{t}} (0) | \hat{t}) \\
                   & = \mathbb{E}[Y_{t}(g) - Y_{\hat{t}} (0) | \hat{t} ] - (\mathbb{E}[ {Y_{t}(0) - Y_{\hat{t}}(0)} | \hat{t} ])
\end{align*}
Now, notice that because we start at the same baseline, we have $Y_{\hat{t}} (0) = Y_{\hat{t}} (g) $. So,
\begin{align*}
    ADS(\hat{t},t) & = \mathbb{E}[Y_{t}(g) - Y_{\hat{t}}(0) \ | \ \hat{t} \ ] - \mathbb{E}[Y_{t}(0) - Y_{\hat{t}}(0) \ | \ \hat{t} \ ] \\
                   & = \mathbb{E}[Y_{t}(g) - Y_{\hat{t}}(g) \ | \ \hat{t} \ ] - \mathbb{E}[Y_{t}(0) - Y_{\hat{t}}(0) \ | \ \hat{t} \ ] \\
                   & = \mathbb{E}[Y_{t}(g) - Y_{\hat{t}}(g) \ | \ \hat{t} \ ] - \mathbb{E}[T_{t} - T_{\hat{t}} \ | \ \hat{t} \ ]
\end{align*}

The above expression is identifiable as all the terms are observed. Hence, using strong parallel trends like assumption and fixing the baseline at some $\hat{t}$, we can estimate ADS.

Our analysis is operationalized through a standard parametric linear regression model that accommodates multi-valued discrete treatment variable, which can be represented by the following regression equation:
\begin{equation*}
    \Delta Z_{i,t,s} = \sum_{t=2009}^{2023} \mathbf{1}\{\tau_i=t\}\gamma_t + \sum_{t=2009}^{2023} \mathbf{1}\{\tau_i=t\}\times\mathbf{1}\{s = student\}\beta_t + \epsilon_{i,t,s}
\end{equation*}

In our study, $\Delta Z_{i,t,s}$ is the change in SAT score for unit i from the baseline year 2008, where $s \in \{student, agent \}$ and $\tau_i$ is the exam year of unit i. The parameter $\beta_t$ is the $ADS(\hat{t},t)$ measuring expected change in high school student's mathematical aptitude relative to baseline 2008 controlling for the exam difficulty through the LLM agent. We additionally leverage parametric MCMC sampling and Bayesian techniques to estimate the $ADS(\hat{t},t)$ for the national level student performance, given we do not have a sample but the mean and standard deviation of the population of SAT taking students through the College Board reports \citep{GelmanBayesian}. A key point requiring some emphasis is that we are not trying to directly compare high school students in 2008 with high school students in some year in the future and claim one to be smarter than the other. This simple comparison between the two is futile whenever we know that the underlying student distribution from the two periods is different. The interpretation of our results are at the discretion of those wish to make such claims. What we are asking though, is that had underlying students behaved in ways similar to LLMs, what would the path of the student SAT scores have looked in that world. This is where the transformed control approach elucidated before allows us to perform this comparison.

3. \textit{Bayesian perspective}. The Bayesian comparison of means strategy offers a nuanced probabilistic approach, enabling us to estimate the full posterior distribution differences in high school students' math abilities across the two distinct periods. This methodology diverges from classical statistical methods by employing Bayesian credible intervals, which provide a more interpretable metric for understanding the probability of observed data under competing hypotheses. In this Bayesian framework, we compute the posterior distributions of the mean SAT scores for each period. The Bayesian credible intervals, in this context, is particularly useful as it quantifies the probability of observing data as extreme as, or more extreme than, the actual observed data, assuming the null hypothesis is true. Unlike traditional confidence intervals, Bayesian credible intervals are derived from the posterior distribution, offering a direct and probabilistic interpretation of the data. The Bayesian credible intervals facilitate a more intuitive understanding of the evidence against the null hypothesis. They provide a direct measure of the probability of the observed outcome under the null, which is often more aligned with the way researchers think about hypothesis testing. This methodological choice is particularly pertinent in our study as it complements the OLS regression analysis, offering a probabilistic interpretation of the changes in students' math performance.

Using the protocol by \cite{Kruschke2013}, we compute the full posterior distributions for the group means before estimating the $ADS(\hat{t},t)$ using the state aggregated and Massachusetts district data. For the national SAT data with population parameters, we employ the Monte Carlo methods by \cite{GelmanBayesian}. The Bayesian estimation procedure transfers credible belief towards the parameter values that are consistent with the data. To begin the procedure, we define straightforward priors given the nature of the exam. All SAT scores for the math section are bounded between \textit{200} and \textit{800}, which we incorporate as the uniform prior distribution for the mean parameters $\mu_{t,s}$ where $t$ is the SAT examination year and the $s \in \{student, agent\}$. These uniformly distributed priors provide necessary bounds on the parameter space while maintaining ignorance over the prior credibility of values within. For the priors on the standard deviation parameters, we follow the proposal by \cite{Gelman2006PriorDistributions}. We used uninformed uniform prior over broad range of values in the strictly positive domain and provide the results in this section. This provides no shrinkage in the estimated standard deviation parameters.

\begin{figure}[ht]
    \centering
    \includegraphics[width=\textwidth,keepaspectratio]{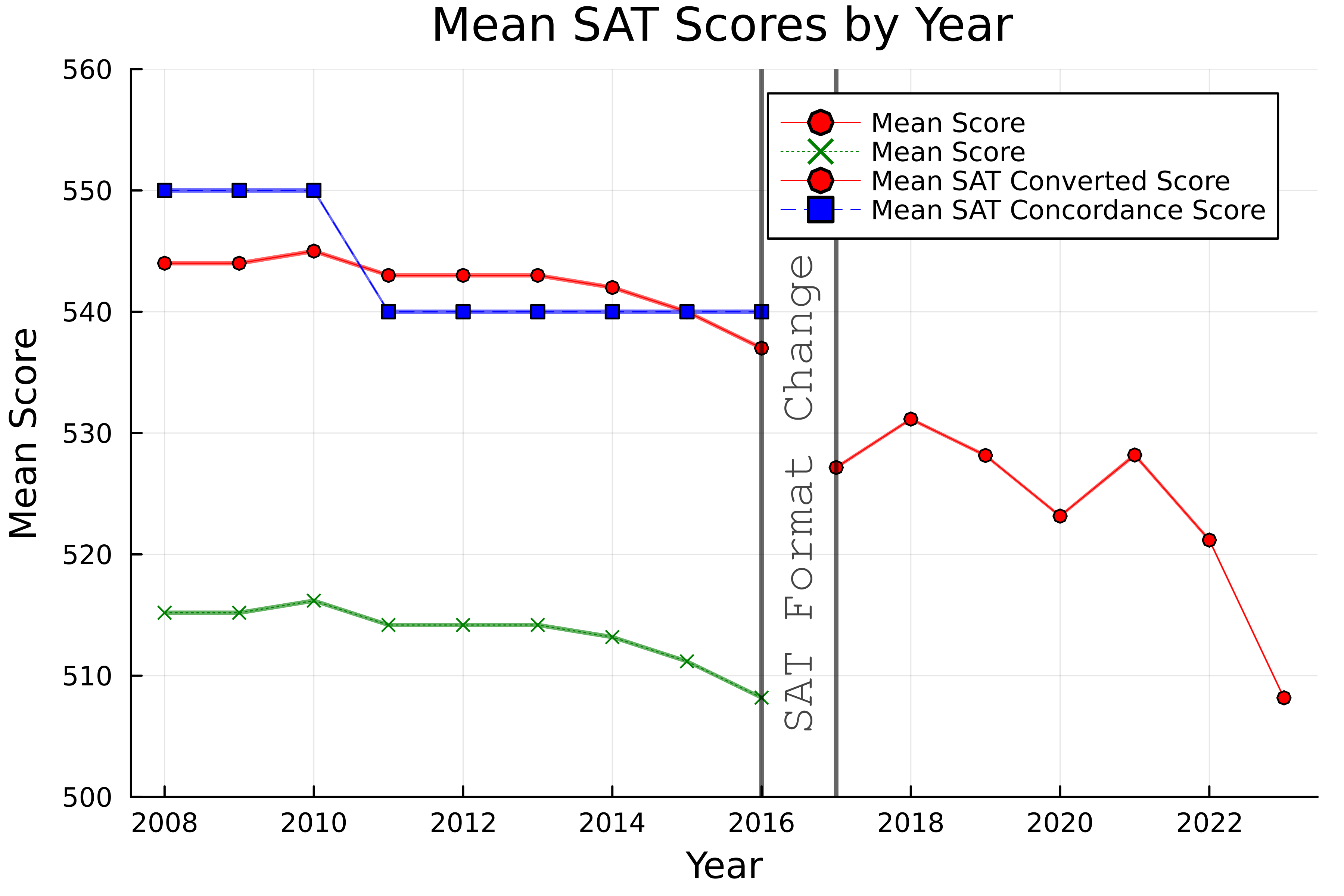}
    \caption{Average SAT Scores of students}
    \label{fig:avg_stud_scores}
\end{figure}

\begin{figure}[ht]
    \centering
    \includegraphics[width=0.6\textwidth,keepaspectratio]{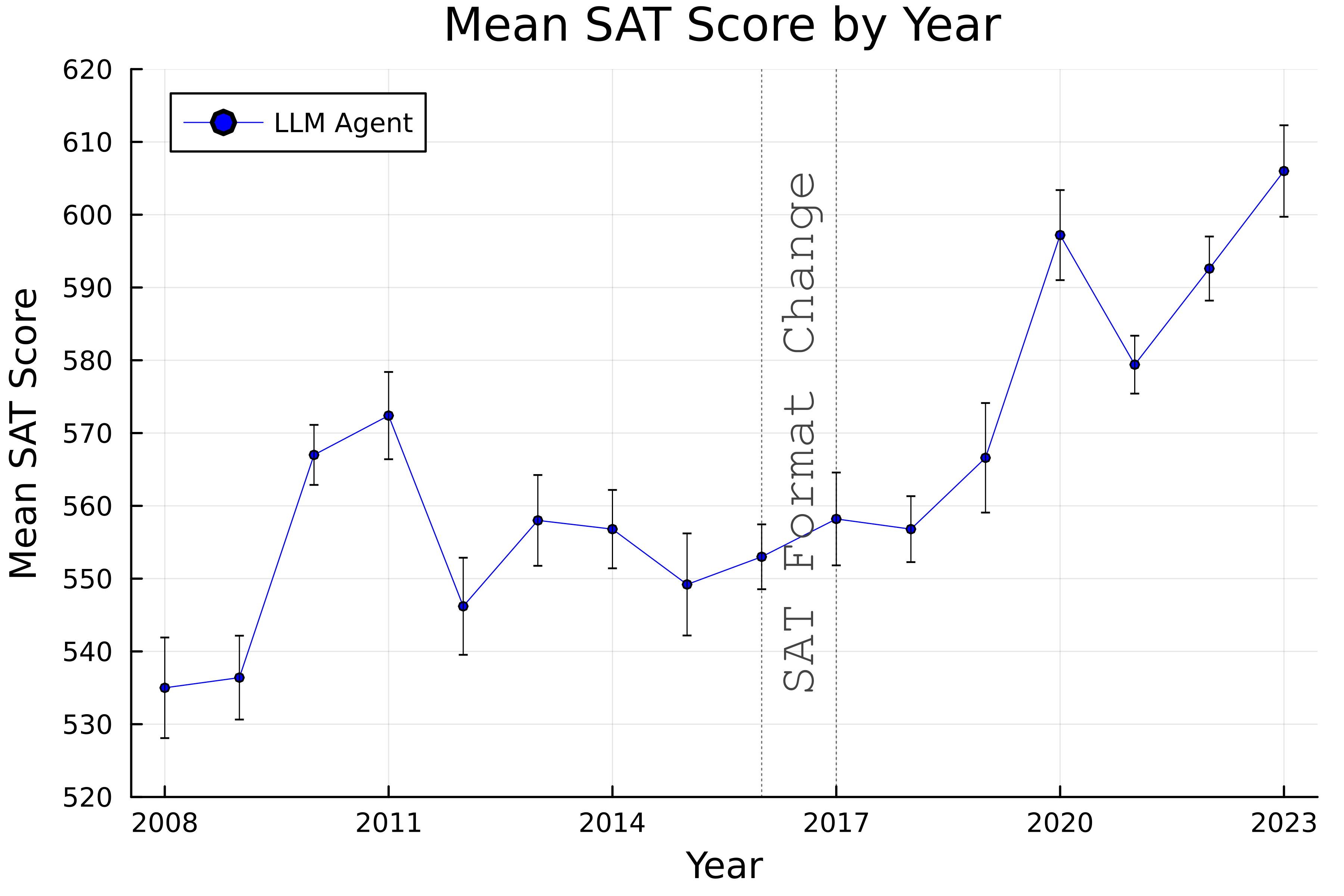}
    \caption{Scaled SAT scores for LLM agent in SAT exams from each year}
    \label{fig:scaled_gpt_score}
\end{figure}

Given the data presented in figure \ref{fig:scaled_gpt_score} and figure \ref{fig:avg_stud_scores}, we would like to establish credible difference in performance for the LLM agent and the students over the periods being considered. For this, we present the Bayesian estimate of the differences within agent and the students. This Normal likelihood is formally presented in the following way:
\begin{equation*}
    Z_{i,t,s} \sim \mathbb{N}(\mu_{t,s},\sigma_{t,s})
\end{equation*}

The posterior sample $\Tilde{\mu}_{t,s}$ for the exam year $t$ and group $s$ is directly employed to infer the posterior $\widetilde{ADS}(\hat{t},t)$ in the following way,
\begin{equation*}
    \widetilde{ADS}(\hat{t},t) = [\Tilde{\mu}_{\hat{t},s} - \Tilde{\mu}_{t,s}|\hat{t}, \text{s=student}]-[\Tilde{\mu}_{\hat{t},s} - \Tilde{\mu}_{t,s}|\hat{t}, \text{s=agent}]
\end{equation*}

\subsection{Scoring methodology}

\begin{figure}[ht]
    \centering
    \includegraphics[width=\textwidth]{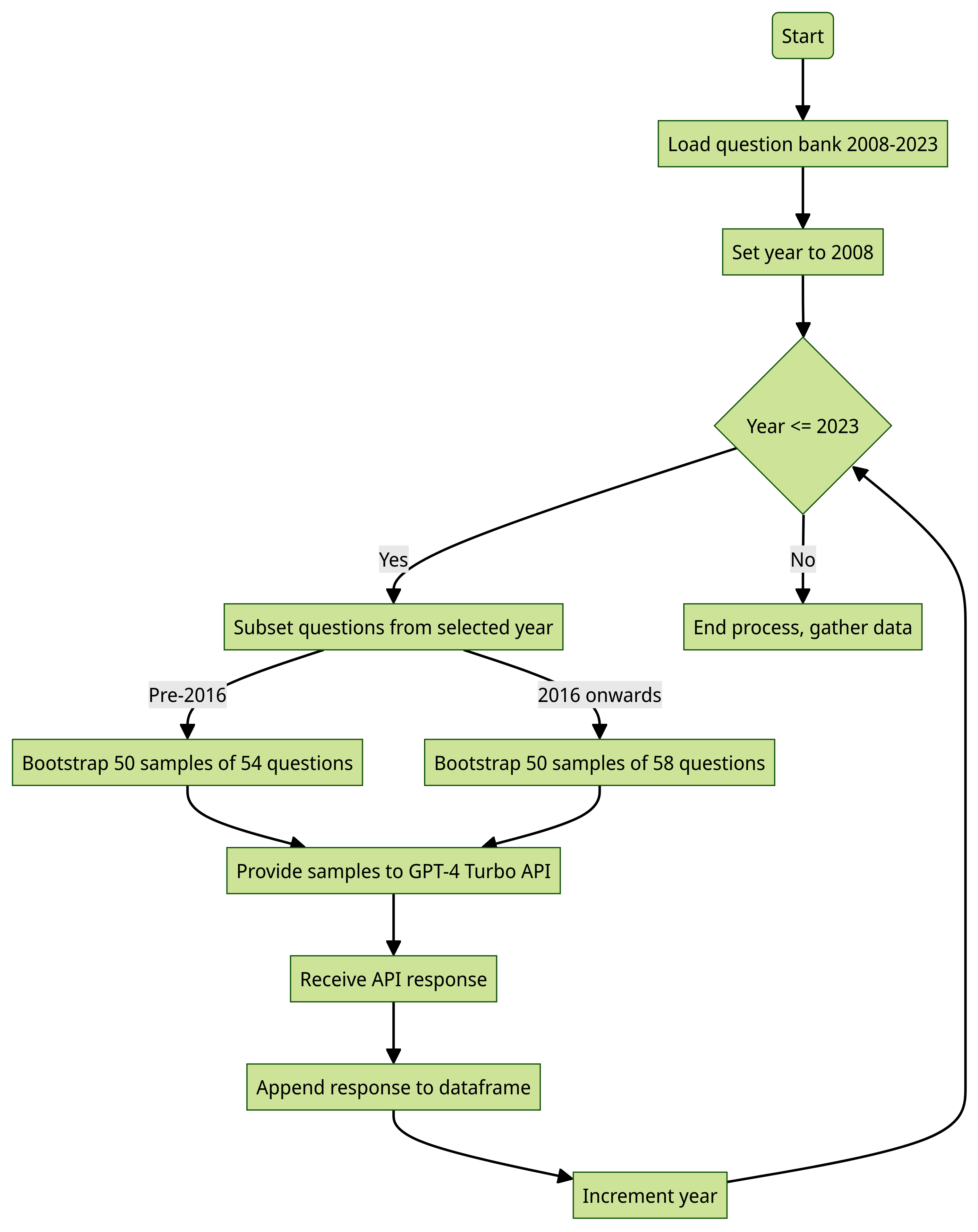}
    \caption{GPT-4 Answering Flowchart}
    \label{fig:gptflowchart_A}
\end{figure}

The section outlines the methodology employed to enable OpenAI's models to answer SAT questions from previous years. From the question bank, we randomly sampled 50 SAT exams for each year using bootstrapping while maintaining identical number of question types as per the SAT format of the given year. We then prompted GPT-4 by OpenAI to take each 50 sampled SAT exams for each year. Then, we calculated the number of correct answers and their proportion (correct divided by total questions) to assess the model's performance and compared the distributions obtained for different periods. In Figure \ref{fig:gptflowchart_A}, we explain how GPT-4 offered by OpenAI answer SAT questions.

\begin{figure}[ht]
    \centering
    \includegraphics[width=0.5\textwidth,keepaspectratio]{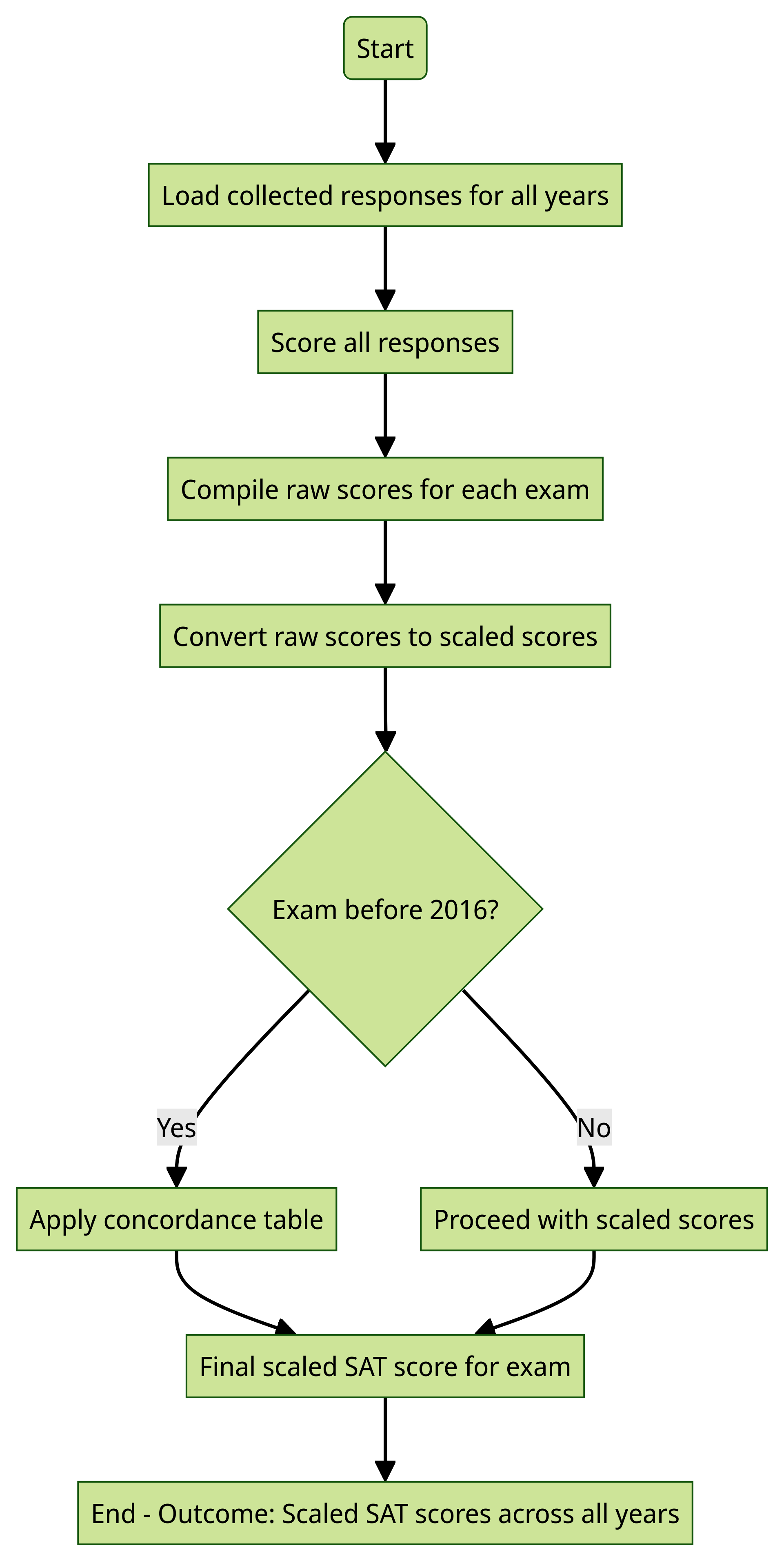}
    \caption{GPT-4 Scoring Process Flowchart}
    \label{fig:gptflowchart_B}
\end{figure}

As illustrated in Figure \ref{fig:gptflowchart_A}, the process of generating answers using the GPT-4 model involves a step-by-step approach. Once all the generated answers are collected, we score this output and make necessary corrections for accurate comparison over time as described in Figure \ref{fig:gptflowchart_B}. First, we sample an SAT exam without replacement from our SAT Math Question bank as explained in section \ref{sec:data}. Next, we proceed by selecting individual questions from the sampled SAT exams, each of which is then presented to the GPT-4 model using a carefully pre-formulated prompt. The structure of our prompt has been provided in the Appendix. Once GPT-4 receives this API call, it provides a response in a format specified by our prompt. The answer is then saved to a dataset that keeps a record of all the questions that have been answered and their corresponding answers. This process continues until the all the questions in the sampled exam have been exhausted. For each period, GPT-4 takes 50 exams, all the questions and answers associated with each exam are compiled into a dataset. The College Board accounts for variation in exam difficulty before scoring the students' performance within that year. As some students are likely to receive relatively more difficult exams than their peers taking the exam in the same year, the final score factors in the difficulty level to allow the accurate assessment of students relative to students in the same peer group. The process of bootstrapping as described generates exams within the year that include questions of varying difficulty, mimicking the student test takers. Thus, some sampled exams are more difficult than others. Thereby, ensuring we evaluate the average SAT exam for a given period.

Formally, $Q$ is our question bank, consisting of questions from each year respectively. $S_t$ is a collection of subsets of $Q$ and each $S_t$ contains randomly sampled exams from the question bank $Q$ without replacement, ensuring unique questions populate each exam. Each exam $E_{j,t} \in S_t \text{ and each } E_{j, t} \subset Q$, where $j = 1, 2, ..., 50$.

\noindent $\forall E_{jt} \in S_{t}, \text{with } t = \{2008, 2009, ..., 2023 \}$ and where $q_{ijt}$ is a question in exam $E_{jt}$, we get

$$n_{mjt} = \sum_{i=1}^{|E_{jt}|} \chi_{A_{q_{ijt}}}(f_m (q_{ijt})) \text{ and } p_{mtj} = \frac{n_{mjt}}{|E_{jt}|} \text{,} $$

\noindent $n_{mjt}$ is the number of questions answered correctly by model $m$, for exam $j$ in period $t$. $p_{mjt}$ is the proportion of questions answered correctly by model $m \in \{\text{GPT-4 Turbo}\}$, for exam $j$ in period $t$. While we restrict this study to GPT-4 model, it can easily be extended to encompass any LLM through open source portals like Huggingface or Ollama. $|E_{jt}|$ is the cardinality of $E_{jt}$, which in our paper represents the number of questions in exam $E_{jt}$. $f_m$ is some LLM agent based function that takes in a question as input and provides an answer as output, depending on whether the answer is for MCQ or answer-type question. $\chi_{A_{q_{ijt}}}$ is a characteristic function that is 1 if output from $f_m$ is correct and 0 otherwise, depending on a set of correct answers $A_{q_{ijt}}$ for question $q_{ijt}$.

\noindent From these performance evaluation measures $n_{mjt}$ and $p_{mjt}$, we obtain the statistics $\mu_{mt}$ and $\nu_{mt}$, along with their respective standard errors $\sigma_{\mu m t}$ and $\sigma_{\nu m t}$ -

$$\mu_{m,t} = \frac{1}{|S_t|} \sum_{j = 1}^{|S_t|} n_{mjt}$$

$$\nu_{m,t} = \frac{1}{|S_t|} \sum_{j = 1}^{|S_t|} p_{mjt}$$

\noindent $\mu_{mt}$ is the mean number of questions answered correctly by model $m$ in period $t$. $\nu_{mt}$ is the mean proportion of questions answered correctly by model $m$ in period $t$. $|S_t|$ is the cardinality of the set $S_t$, which is 50 for all time periods.

\subsection{Prompting}
OpenAI provides access to their GPT series models through Application Programming Interface (API) endpoint. In this study, we used the most advanced model available GPT-4 Turbo. This model has been bench marked by OpenAI to be performing at the 89th percentile on SAT mathematics under certain prompting conditions as seen in \cite{openai-gpt4}. For the purposes of this study, we employed a prompting strategy with explicit system level instructions to return the appropriate character letter corresponding to the correct answer having analyzed the question provided. Prompting strategies can affect the performance of the model as described by \cite{gpt-brazil2023}. Since, we are concerned with change in the LLM agent performance over time with reference to baseline, any idiosyncratic LLM effects and prompt effects cancel out. We employed a zero-shot prompt, by which, no examples solutions were provided to the model to facilitate answering the question. Further, we did not allow for chain-of-thought. The model was proscribed from sequentially reasons itself to the appropriate answer. The question was provided through the prompt and the model was asked to provide one character letter output for multiple choice questions and the appropriate numerical or equation output for the answer type questions. The prompting strategy is identical over all periods, ensuring there is no bias through the prompt. Since all prompts are independent API calls to the model, the model has no memory of previous questions. The model therefore is tasked to answer each question independent of any other question or reference to the period from which the question is gathered. This ensured the comprehensive evaluation of the difficulty of the questions from appropriate periods. Further prompting details and GPT-4 model parameters are provided in the appendix.

\section{Results} \label{sec:results}

First, we share the performance of GPT-4 on the exams from different years. We use these results as a benchmark to then estimate the change in high-school student performance. We estimate the change in student performance at the national level and in the state of Massachusetts. We also provide the results for different demographic groups to understand the heterogeneity in the change in student performance over time.

\subsection{GPT-4 taking the SAT exam}
\begin{figure}[ht]
    \centering
    \includegraphics[width=0.6\textwidth,keepaspectratio]{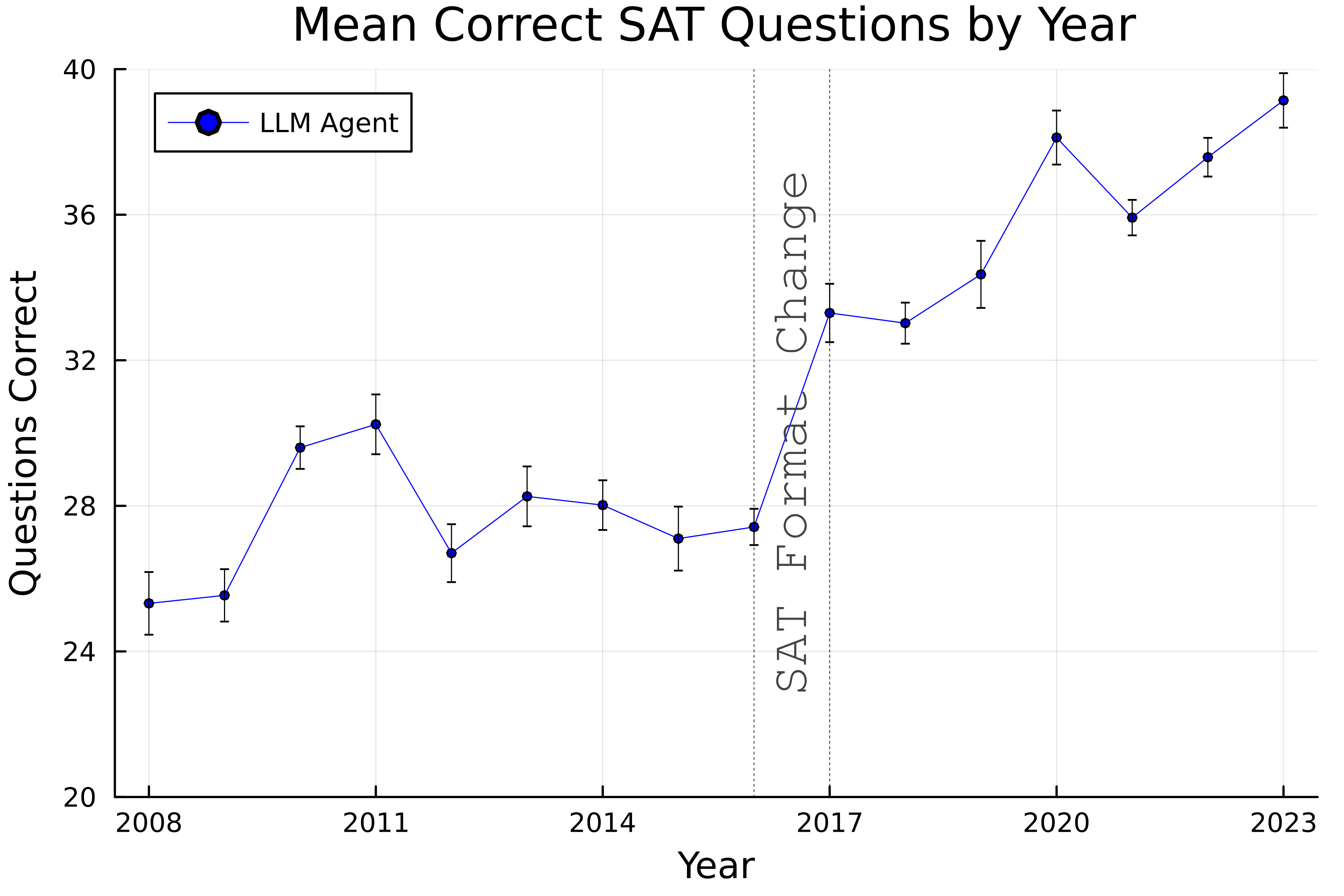}
    \caption{Raw correct answers by LLM agent in SATs from each year}
    \label{fig:raw_gpt_score}
\end{figure}

The Figure \ref{fig:raw_gpt_score} shows the number of SAT questions GPT-4 answered correctly in SATs from a given year. The GPT-4 model's average performance over time is seen to be increasing and especially after the format change. As there are 4 additional questions after the format change, it is challenging to interpret the raw number of questions answered by GPT-4 before and after the format change. The Figure \ref{fig:scaled_gpt_score} provides the corresponding scaled SAT score which can be interpreted. We use the concordance tables provided by the College Board to recenter the scores before the format change as described in Section \ref{sec:data} with the appropriate conversion. Given the scaled SAT score can be compared over time, the performance of the GPT-4 model can be noted to steadily increase over time with reference to the baseline year 2008. As highlighted in Section \ref{sec:data}, this difference is the change in difficulty of the exam, and not due to syllabus change.





In evaluating GPT-4's performance on SAT math sections, we accounted for the varying number of questions in the \textit{pre} format change (54 questions) and \textit{post} format change (58 questions) periods. To enable a fair comparison, we also analyzed the proportion of correctly answered questions for each year. i.e, the ratio of number of correct answers to total questions per exam over 50 bootstrap samples. This approach ensures a normalized comparison despite the different total question counts. Our analysis revealed consistent results with our initial findings. This proportional comparison aligns with our earlier observations, indicating a noticeable decline in the rigor of SAT math sections in the \textit{post} period as the LLM agent answers a higher proportion of questions accurately.

\subsection{GPT-4 vs High School Students}

1. \textit{Mean Comparison}. We begin by providing results for a straightforward estimation of the $ADS(\hat{t},t)$ for the baseline year 2008. This gives us the relative change in the performance of average high school student, having controlled for the SAT exam difficulty using the transformed control. We perform this comparison at the national level and for the state of Massachusetts. Our state level comparison only comprises of school districts within the state of Massachusetts.

\begin{figure}[htbp]
    \centering
    \includegraphics[width=\textwidth,keepaspectratio]{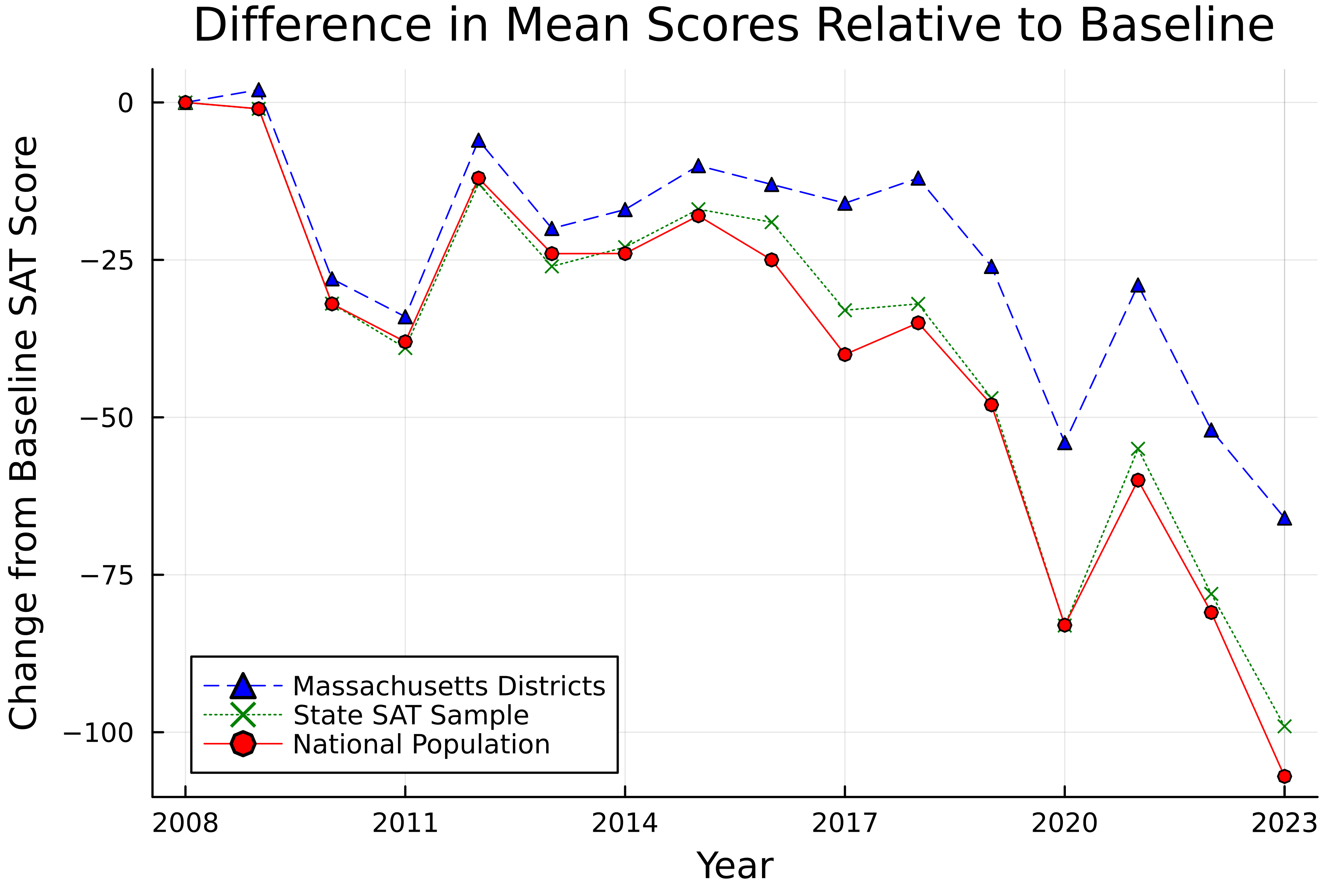}
    \caption{Estimated $ADS(\hat{t},t)$ with reference to baseline year 2008}
    \label{fig:results_mean}
\end{figure}

The Figure \ref{fig:results_mean} shows the mean difference in SAT scores between the average high school student and the transformed control group, relative to the baseline year of 2008. All three levels of comparison show a downward trend in the comparison, and the trend highlights declining students' mathematical ability over the years. It is important to note the sample of unweighted state SAT score sample tracks and mimics the national level decline.

Table \ref{tab:yearly_sat_decline} provides a comprehensive overview of the annual changes in SAT scores at the national using aggregated state SAT and Massachusetts levels, from 2009 to 2023. Each year's figures are benchmarked against the scores from 2008, thereby offering a year-over-year comparative analysis of performance across different geographical tiers.



These results highlight key trends; state aggregated data generally mirrors national trends but with varying degrees of change. The sharp decline in 2020 and again in 2023 might point to statewide educational challenges or changes in state-specific educational policies or practices. Massachusetts presents a slightly different narrative, with changes in SAT scores that often align with national trends but with some exceptions. For instance, the decline in Massachusetts is somewhat less pronounced in certain years, indicating possible state-specific factors that might have cushioned the impact of broader educational challenges. Overall, we find a statistically significant difference in mathematical performance of the agent and the average performance of high school students, especially in more recent years.

2. \textit{Bayesian comparison}. We estimate the $ADS(\hat{t},t)$ by first estimating the posterior distributions using the methods provided by \cite{GelmanBayesian} and \cite{Kruschke2013} to compute the difference in the means between the appropriate groups.

We include individual parameters for standard deviation within the groups through $\sigma_{t,s}$. The Bayesian approach can seamlessly incorporate unequal variances within the groups and using the Bayes Rule we can reallocate belief given the data to arrive at the posterior distribution. To obtain marginal posterior distributions for the parameters of interest, we employ the Hamiltonian Monte Carlo (HMC) method described by \cite{Betancourt2017}. Since the Bayesian approach provides full posterior distribution of the credible values of the parameters, we apply the standard rejection rule of 5\% p-value to reject the null value. The estimated $ADS(\hat{t},t)$ at the national level using population parameters and unweighted SAT state averages along with school district level for Massachusetts are provided in Table \ref{tab:bayes_sat_decline}. The posterior mean estimate can be noted to be statistically identical to results in Table \ref{tab:yearly_sat_decline}. While it is possible to naively infer changes in student performance over time, we provide estimates for difference in performance that far larger than previously anticipated through the ‘transformed control' paradigm. This reveals an alarming 107 point decline in student performance in the SAT math section compared to the baseline year 2008.

\section{Heterogeneity Analysis} \label{sec:het}

\begin{figure}
    \centering
    \includegraphics[width=\textwidth,keepaspectratio]{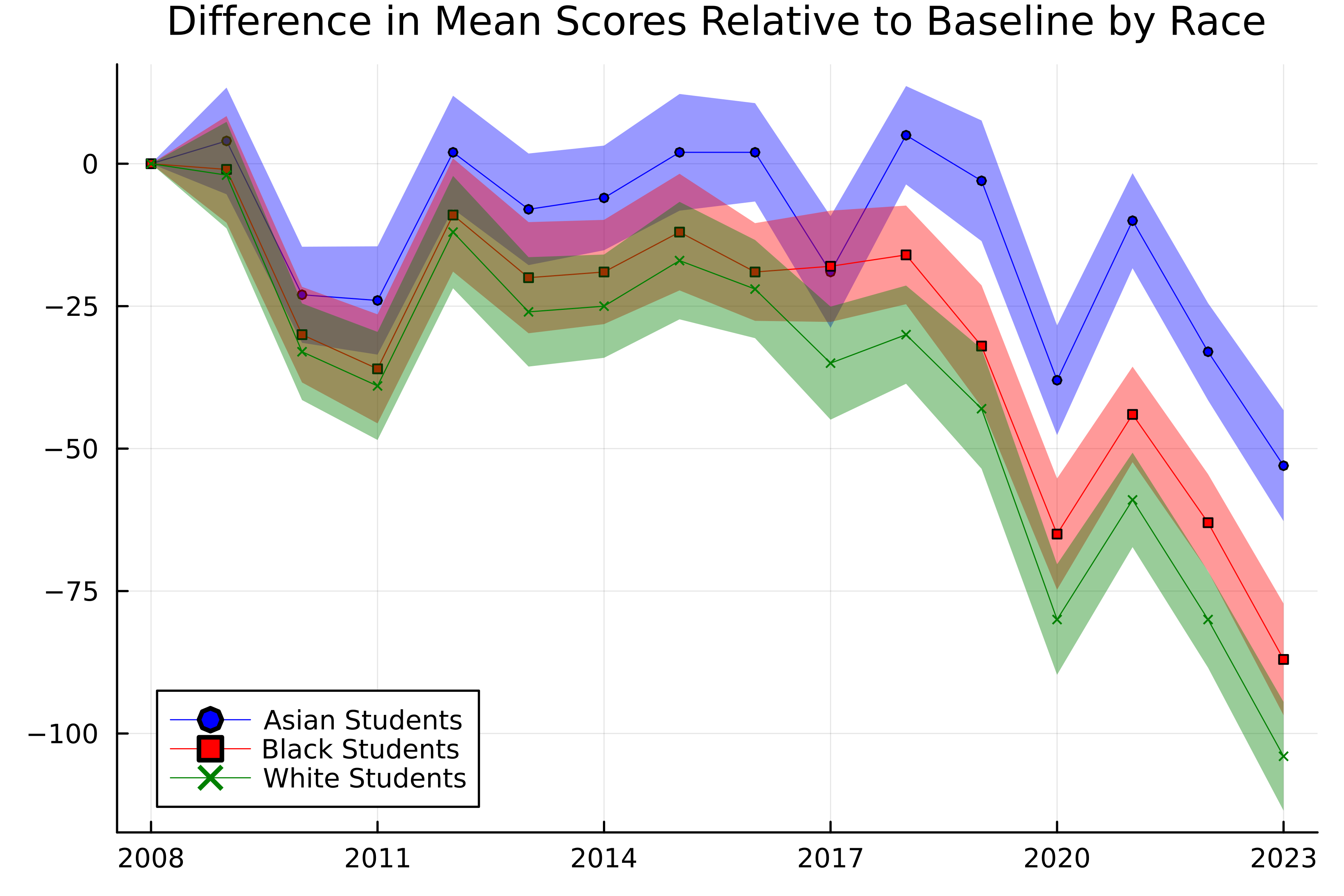}
    \caption{Estimated ADS for Asian, Black and White Students}
    \label{fig:race_ads}
\end{figure}

In this section, we identify the $ADS(\hat{t},t)$ parameter across a few demographics. The ADS as described provides a richer interpretation of the difference in student performance over time when controlling for the difficulty of the exam. Since students from different demographics take the same standardized test in a given year, their performance can be compared to that of students from the same demographic group who took the test in a different year. The estimated $ADS(\hat{t},t)$ of the Asian, Black and White students is shown in Table \ref{tab:race_ads}. We employ the Bayesian approach described in the earlier sections to arrive at the full posterior distribution of the $ADS(\hat{t},t)$. White students show the largest difference in performance of 104 SAT points from baseline year. The Figure \ref{fig:race_ads} shows the steep decline in performance in recent years. The College Board reports do no have consistent demographic information over the period considered for Hispanic students, we therefore considered the most consistent demographic breakdown of the SAT scores present in the report. Similarly, male students show a decline of 117 points while female students decline by 100 points.





\section{Mechanisms} \label{sec:mech}

\cite{hoxby2009} sets the premise to understand the declining rigor of the SAT exam we observe. First, \cite{hoxby2009} defines selectivity of universities based on the percentile rank of admitted students in the standardized tests. From 1962, the elite plus universities (see \cite{Chetty2023-mf}) remained very competitive, selecting the top percentile students in the standardized tests. Whereas, the absolute standard of achievement required of a freshman who successfully competed for a seat at less selective universities was falling from 1962. Colleges correspondingly responded by endogenously adapting to the new student population. It is likely that the College Board similarly has been endogenously changing the SAT to match the demands of the less selective universities. After all, as \cite{hoxby2009} notes, a 100 point SAT math score range 700-800 corresponds to less than 10 percentile of student performers (admitted to high selective universities), while a 450-550 score range corresponds to 33 percentile of student performers (admitted to median selective universities). The exam still discriminates the top percentile of students while providing a respectable score for the median student entering a less selective university.

Understanding the decline in math performance among high school students, as evidenced by our analysis of the SAT math section difficulty from 2008 to 2023, necessitates an exploration of the potential underlying mechanisms. These mechanisms help us elucidate the broader socio-cultural, political, technological and educational factors that impact the mathematical performance. Herein, we delve into four plausible channels that could contribute to the observed decline.

\textbf{1. Test Optional Policy:} The increasing prevalence of test-optional policies among universities has potentially reduced the emphasis on standardized testing, leading to a decline in student motivation and preparedness for the SAT math section. To further understand this mechanism, we conducted a college-wide survey to gauge students' and faculty's perceptions of the SAT and its importance in the admissions process. Our survey\footnote{For further details related to the survey, see the Appendix \ref{sec:appxf}.} revealed a significant divergence between students' and faculty's perception about the role of SAT in college admissions process and post-college outcomes. Surveyed faculty believe students admitted in 2008 would have had 33-point higher SAT score than students admitted in 2023. On the other hand, students did not consider SAT scores as important for college admissions, with many viewing it as less critical for college admissions and post-college outcomes. The reduced importance of the SAT in college admissions could have inadvertently impacted students' engagement with mathematical learning, contributing to the observed decline in math performance.

\textbf{2. Increased Screen Time}:  A substantial body of research posits that the heightened time students devote to their cellphones and other digital devices adversely impacts their cognitive and academic performance. Jonathan Haidt, among others, has extensively documented the correlation between increased screen time and diminished attention spans, directly affecting students' ability to engage with complex mathematical problems \citep{haidt2023get}. The pervasive presence of digital technology not only diverts attention from academic pursuits but also potentially reduces the time available for focused study and problem-solving, essential for developing mathematical proficiency. This mechanism is consistent with our findings, where the decline in SAT math scores over time may be partly attributable to students' fragmented attention and decreased engagement with mathematical learning due to excessive screen time.

\textbf{3. Grade Inflation}: Grade inflation emerges as another possible mechanism, complicating the landscape of academic assessment and student performance evaluation. This phenomenon, where grades increase over time without a corresponding rise in learning and ability, might create a misleading portrayal of student capabilities. According to \cite{sonner2000adjunct}, grade inflation can lead to student complacency, potentially causing students to underestimate the rigor of standardized tests like the SAT math section. The disconnect between perceived and actual proficiency could explain part of the decline observed in our study, suggesting that students may not be as prepared for the SAT's mathematical challenges as their classroom grades might indicate.

\textbf{4. Mental Health Decline}: Lastly, the decline in students' math performance may also be tied to the broader issue of declining mental health among adolescents and young adults. Significant literature underscores the negative impacts of mental health issues, such as anxiety and depression, on academic performance \citep{twenge2019age}. These conditions can drastically affect students' concentration, motivation, and overall academic performance, including on standardized tests like the SAT. The additional stress associated with high-stakes testing can exacerbate these issues, leading to outcomes we observed. Our findings, highlighting a steeper decline in student proficiency than the softening of SAT math question rigor alone could account for, suggest that external factors such as mental health could be significantly influencing student performance.

\section{Conclusion} \label{sec:conclusion}

This paper contributes to the discourse on standardized testing and educational assessment, revealing a decrease in SAT math rigor over time. Employing OpenAI's GPT-4 model through the ‘Transformed Control' paradigm, we present a novel approach to construct counterfactual data. This circumvents the limitations of traditional control groups and offers a nuanced understanding of test difficulty over time. Our findings underscore a dual trend: a discernible decrease in the rigor of SAT math from 2008 to 2023, accompanied by a more pronounced decline in students' math performance over the same period. This raises important questions about the SAT's role as a tool for university admissions, suggesting that the test may not be as challenging as it once was, potentially diminishing its effectiveness in distinguishing among applicants based on their mathematical proficiency.

Further, the decline in student performance highlights broader educational challenges, prompting a reevaluation of teaching strategies, curriculum design, and the overall approach to mathematical education in high schools. Stakeholders in the educational sector—educators, policymakers, and institutions—must consider these trends in their efforts to foster a robust mathematical foundation for students. This research also demonstrates the potential of leveraging AI and machine learning models to enrich social science research, providing a template for future studies in educational assessment and beyond.


\pagebreak

\bibliography{qje_references}  

\pagebreak

\begin{appendix}
    \renewcommand{\thetable}{\Alph{section}.\arabic{table}}

    \section*{Appendix A: Concordance Tables} \label{sec:appxa}

    The concordance tables provided below come directly from the College Board. We utilize these concordance tables in our paper for 2 types of score conversions: converting raw to scaled score and converting old to new score. The table below shows the conversion tables utilized by our study.

    \section*{Appendix B: Data Sources} \label{sec:appxb}

    After collecting the PDF files, we transcribed each SAT exam PDF into a structured comma-separated values (CSV) file. MCQs present a set of possible answers, requiring the examinee to select the most appropriate option. Answer type allowed a one-line input from the examinee.

    The College Board provides yearly reports with the population level SAT performance by the cohort of high school students taking the SAT exam. The data from these reports include average scores, total test takers and standard deviation in the scores for the mathematics test and the language and writing test. The concordance table provided by the College board requires rounding the average SAT score for exams in the \textit{pre} period to the nearest multiple of 10 before it can be mapped to the average SAT score based on the concordance table. Due to this, the direct conversion of SAT scores using the concordance table has certain limitations. For example, if an average SAT score is 514 in the year 2009 and the score is 515 in the year 2010, 514 would be rounded to 510 and 515 would be rounded to 520, before the concordance table can be used. To avoid this problem, we used a simple linear regression model to regress average SAT scores before and after the conversion in the concordance table and then linearly interpolate the average SAT scores, which we call the Concordance SAT scores.

    \subsection*{Negative marking}
    As previously stated, our study spans two eras, 2008-2016 known as \textit{pre} and 2017-2023 known as \textit{post} period. During \textit{pre} period, the exams implemented negative marking, unlike the 2017-2023 period. In analyzing the GPT models' performance, we opted not to apply negative scoring for incorrect responses when converting from raw score to scaled score. This decision might initially suggest an inflated assessment of GPT performance in the \textit{pre} period, as measured by scaled SAT scores. However, as our forthcoming sections demonstrate, incorporating negative marking for this period would likely have amplified the performance discrepancy between the two periods. Therefore, our analysis likely presents a conservative estimate, potentially downplaying the actual performance gap.

    \subsection*{Syllabus change}
    Although SAT math section did not experience a major change in structure and syllabus as compared to the verbal section, we must emphasize certain differences in exam content between the two periods. Before 2016, the key focus areas were arithmetic, numbers and operations, algebra, functions, geometry and data analysis. Starting 2016, SAT exam expanded the range of topics further and included questions related to trigonometry and complex numbers. Also, there was a greater focus on data analysis, graphs and word problems and less emphasis on geometry-related questions. Furthermore, starting 2016, students started received sub-scores for sections labeled as \textit{Heart of Algebra}, \textit{Passport to Advanced math} and \textit{Problem Solving and Data Analysis}. There were no such subsections before 2016. Just by looking at the syllabus, one could argue that the math section became harder than before as it covered the number of topics covered increased, along with their difficulty level.

    \subsection*{Data Sources}
    These reports provide comprehensive insights into the SAT performance of students on a state-by-state basis, enabling a detailed examination of trends and patterns in SAT scores across the United States. However, the availability of state-level SAT data prior to 2016 posed a unique challenge. To procure this historical data for the \textit{pre} period (2008-2016), we used the Internet Archive and the National Center for Education Statistics. This digital archive, renowned for its extensive collection of web pages archived over time, proved instrumental in retrieving past state-level SAT reports not readily accessible on the College Board's current website. Utilizing this, we systematically sourced and compiled state-level SAT score reports for each year from 2008 to 2016. This approach of combining current and archived data sources ensures a comprehensive and continuous dataset spanning the entire duration of our study period.

    \subsection*{Data Cleaning and Recentering}
    The distribution of the average scores for these school districts is provided in Figures \ref{fig:avg_stud_scores_mass} and \ref{fig:avg_stud_scores_mass_conv}, which show the distribution of average scores before and after the recentering using the concordance tables from the \textit{pre} period to match the \textit{post} format change period.


    \section*{Appendix C: Prompting} \label{sec:appxc}

    OpenAI provides access to their GPT series models through Application Programming Interface (API) endpoint. In this study, we used the most advanced model available GPT-4 . This model has been bench marked by OpenAI to be performing at the 89th percentile on SAT mathematics under certain prompting conditions as seen in \cite{openai-gpt4}. For the purposes of this study, we employed a prompting strategy with explicit system level instructions to return the appropriate character letter corresponding to the correct answer having analyzed the question provided. Prompting strategies can affect the performance of the model as described by \cite{gpt-brazil2023}. Since, we are concerned with change in the LLM agent performance over time with reference to baseline, any idiosyncratic LLM effects and prompt effects cancel out. We employed a zero-shot prompt, by which, no examples solutions were provided to the model to facilitate answering the question. Further, we did not allow for chain-of-thought. The model was proscribed from sequentially reasons itself to the appropriate answer. The question was provided through the prompt and the model was asked to provide one character letter output for multiple choice questions and the appropriate numerical or equation output for the answer type questions. The prompting strategy is identical over all periods, ensuring there is no bias through the prompt. Since all prompts are independent API calls to the model, the model has no memory of previous questions. The model therefore is tasked to answer each question independent of any other question or reference to the period from which the question is gathered. This ensured the comprehensive evaluation of the difficulty of the questions from appropriate periods.

    The important parameters concerning this process are the temperature of the model and maximum token output limit. The temperature of the model governs the randomness of the output, the parameter when set to 0 provides a consistent and unwavering output from the LLM agent. This ensures the output from the LLM agent is deterministic and repetitive by always selecting the most probable token given the previous token. As the temperature is increased to higher levels, the agent has additional creative license to provide more abstract outputs. Since all LLM agents are predictive engines for the next token, higher temperature increases the probability of less likely token being selected as the output. For the purposes of our experiment and not to induce bias through the prompt temperature, we set it to 0. The maximum token output parameter limits the LLM agent's output. Since the agent in our experiment is required to choose the correct multiple choice option or provide straightforward numerical output, we limit the agent to 5 tokens, which is roughly 15-20 characters.

    Our prompt template is shown in Listing \ref{lst:sat_instructions}, it encapsulates the strategy employed throughout the experiment. The LLM agent operates as an assistant to the user and to illicit a response from this assistant, two level of prompts are used. The system-level prompt is a high level instruction that the assistant is expected to follow verboten, this gives necessary context to the LLM agent and appropriately modulates its responses. For example, the LLM agent can be required to respond as a Shakespearean character through the system-level instructions. This is less useful in our case but provides the necessary platform to provide broad instructions to the agent about the task it is undertaking. The user-level prompt then provides the SAT question to the agent, the agent then responds to the output based on the temperature and maximum token output limit. Once these prompts and model parameters are held constant, we loop through the bootstrapped exams from each year as shown in Figure \ref{fig:gptflowchart_B}. This ensures our agent is neutrally evaluating the SAT exam in the selected year while all other parameters are held constant.

    \begin{lstlisting}[caption={SAT math Exam Instructions for GPT-4}, label={lst:sat_instructions}, breaklines=true]
    @System-level prompt@
    You are taking an SAT math exam which include multiple choice and answer type questions.
    Determine the correct answer.
    Choose only ONE 'character letter' response output corresponding to the correct answer from the options provided for multiple choice type.
    Provide the appropriate numerical answer as required for the answer type question without any units of measurement.

    @User-level prompt (for multiple-choice questions)@
    You are provided with an SAT question enclosed in triple backticks, followed by multiple choice options.
    ``` <Question> 
        <Options> ```
    Please identify and return ONLY ONE letter character corresponding to the correct option.
    Your response output should only be ONE character letter.

    @User-level prompt (for numerical answer questions)@
    You are provided with an SAT question of numerical type, enclosed in triple backticks. Please determine and return the correct numerical value or mathematical expression.
    ``` <Question> ```
    WARNING: Do not provide any explanations, calculations, units of measurement, or additional outputs.
\end{lstlisting}

    \section*{Appendix D: Other LLMs} \label{sec:appxd}

    To ensure the robustness of our findings and to validate that our results are not subject to the specific LLM used, we conducted additional tests using different versions of the LLM. We utilized the GPT-4 April update and another LLM, Claude 3.5 Sonnet, for this purpose. 

    GPT-4 Turbo received an update on April 9th, 2024. This update by OpenAI majorly improved the model performance \citep{openai2023gpt4turbo}. The results in the main text are from GPT-4 January update, and we provide the results from GPT-4 April update in Table \ref{tab:bayes_sat_decline_gpt4}. The results show a similar trend in the decline of student performance over time. The decline in SAT scores is 113 points at the national level in 2023 compared to 2008. The decline in SAT scores is 72 points at the Massachusetts level in 2023 compared to 2008. Additionally, it can be noted through Figure \ref{fig:gpt4_april_jan_claude_diff} that the performance of GPT-4 April tracks the GPT-4 January in evaluating underling SAT. The results are consistent with the main text and show a decline in SAT math rigor over time. 

    Next, we utilized Claude 3.5 Sonnet, a different LLM, to evaluate the SAT exams. Claude 3.5 Sonnet is a LLM developed by Anthropic, a U.S based artifical intelligence (AI) company. We used this model as model benchmarking performed by Anthropic show similar performance results compared to OpenAI's GPT-40 model \citep{anthropic2024}. The results from Claude 3.5 Sonnet are also provided in Figure \ref{fig:gpt4_april_jan_claude_diff}. In the figure, we can clearly see that the initial and end point performance of Claude 3.5 Sonnet on the SAT exams is consistent with the performance of GPT-4 January and GPT-4 April update.

    The overall trend of declining SAT scores from 2008 to 2023 was replicated by these runs, further strengthening our claim. This robustness test demonstrates that our findings are consistent across different LLMs, provided that the intelligence demonstrated by LLMs is comparable, and are not an artifact of the specific LLM used in the initial analysis.

    \section*{Appendix E: Additional Tables and Figures} \label{sec:appxe}

\begin{figure}[ht]
    \centering
    \includegraphics[width=\textwidth,keepaspectratio]{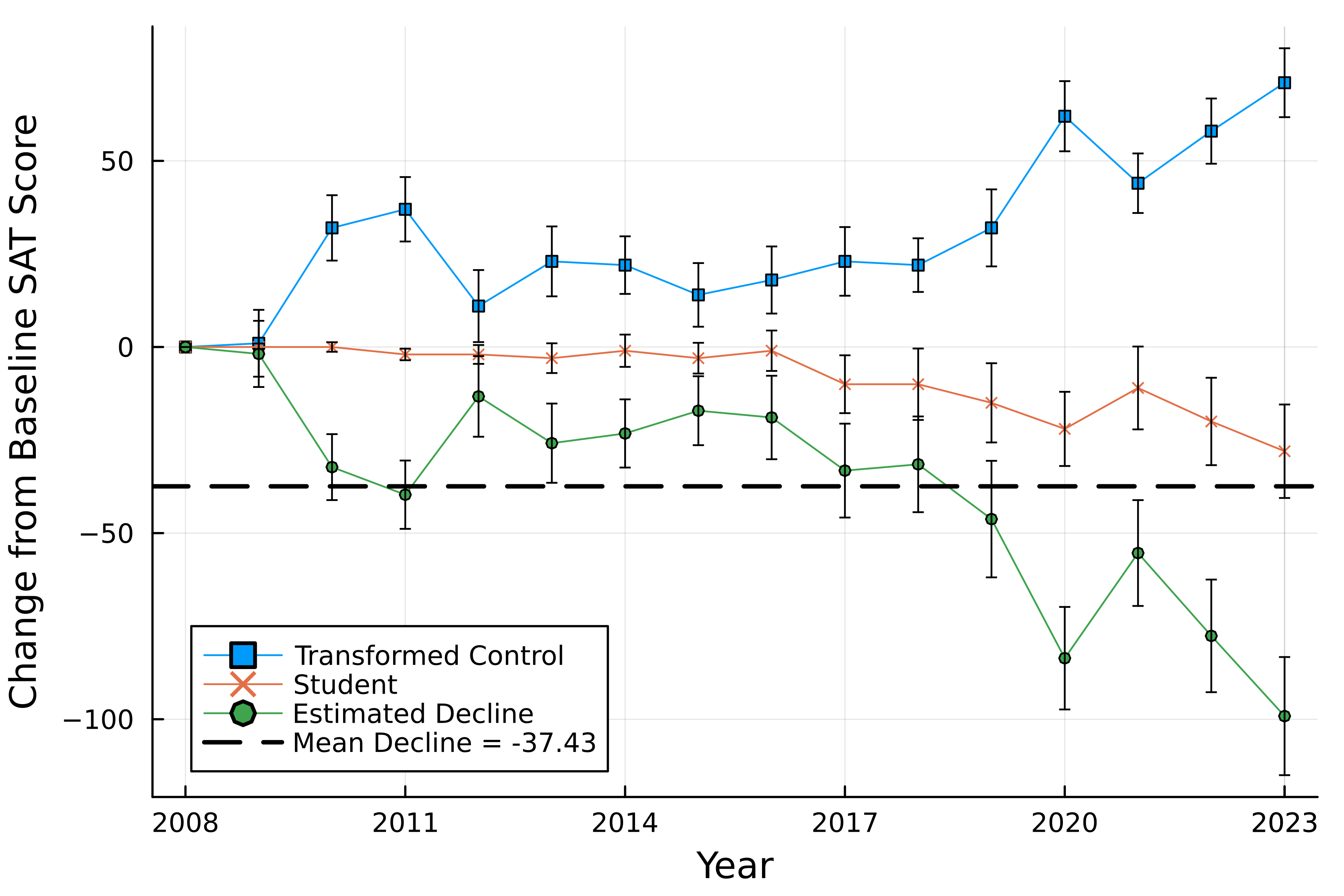}
    \caption{Bayesian Estimated $ADS(\hat{t},t)$ using unweighted State SAT averages}
    \label{fig:bayes_state_nopool}
\end{figure}

    In Figure \ref{fig:bayes_state_nopool} and Figure \ref{fig:bayes_mass_nopool} we include the posterior mean and 95\% credible interval for relative performance of the LLM agent with reference to baseline 2008 score in the SAT exam over time. Similar posterior estimates for relative student performance using the state SAT data and Massachusetts school district data are included in orange. The posterior estimates for $ADS(\hat{t},t)$ shine light on a dramatic difference in student performance having controlled for the change in SAT difficulty over time. The average decline in student performance is 37 points over the period, while the decline at the endpoint in 2023 is 99 points compared to 2008 using the unweighted state SAT data. The Figure \ref{fig:results_mean} shows the unweighted state SAT data traces the national trend.

    A majority of the universities according to \cite{FairTest} made SAT score submissions optional during the Covid-19 pandemic for fall 2021 admissions. For this reason, we removed SAT exam results for 2021 and onwards as students may have been less prepared knowing the universities are not emphasizing the SAT. The results remain robust to exclusion of COVID years, and the estimates did not show any significant difference. The difference in high school student performance in the SAT is 84 points lower compared to the baseline 2008 score at the national level in 2020.

    \pagebreak

    \section*{Appendix F: Lindner College of Business Survey} \label{sec:appxf}

    We conducted a general attitude survey about standardized testing for college admissions among undergraduate students (N=194) and faculty (N=42) at the Lindner College of Business, University of Cincinnati\footnote{On 4/17/2024, the IRB reviewed the submission and determined that this protocol meets the criteria for exemption from IRB review in accordance with 45 CFR 46.104.}. The University of Cincinnati has been a test optional institution since 2021. While these are a small samples, they provide valuable insight about how the test is perceived at a medium sized state university. The notable points from the survey; students underestimate the value of the SAT and it's predictive power, while the faculty believe the SAT can predict student performance and post-college outcomes. The faculty also note the the average SAT math score of a student admitted to the Lindner College of Business in 2008 would have been higher compared to student admitted in 2023. This mirrors our results and lends credence to notion of universities becoming less selective and endogenously updating their standards to accommodate new students \citep{hoxby2009}. The College Board could therefore be reacting to the updated demands of universities by weakening the standards of the SAT math. The following figures summarize the key findings from survey, the questions are motivated by \cite{Chetty2023-mf}.

    \section*{Appendix G: Question Embeddings and Difficulty Alignment} \label{sec:appxg}



    The College Board in exam years between 2008 and 2014 provided difficulty rating for 780 collected questions during this period. The difficulty rating is a score between 1 and 5, with 1 being the easiest and 5 being the hardest. We use the rating to ensure the LLM agent's performance is aligned with the difficulty of the questions. For this task, we use the text-embedding-3-large model from OpenAI \citep{neelakantan2022text}. Embeddings are a way to represent text in a numerical format, which contains the semantic meaning of the text. These embeddings show how the LLM agent identifies the text in its internal representation.

    The embeddings created by the model are used as features to predict the difficulty of the questions. In addition, the SAT is a paper and pencil test for the years considered in this study, and the difficult questions appear at the end of each section. We use this information as an additional feature in the classification model by including a progress bar as a numerical representation of the progress through the section. To simplify the classification task, we group difficulty ratings 1 and 2 as easy (310 questions), 3 as medium (265 questions), and 4 and 5 as hard (205 questions). The model is trained on 80\% of the data having balanced the classes and tested on the remaining data. Using a Random Forest Classifier, the model achieves an accuracy of 0.789 on the testing data. The confusion matrix for the classification accuracy is shown in Table \ref{table:confusion_matrix}. We tune the parameters of the classifier to achieve the accuracy and generalize to the unseen data. The model is then used to predict the difficulty of the questions in the SAT exams from 2015 to 2023.

    The results are shown in Figure \ref{fig:difficulty_rating}, which highlights the proportion of easy, medium, and hard questions in the bootstrapped exams from each year. Further, the average proportion of easy, medium, and hard questions in the complete SATs gathered for the years 2008 to 2014 is 0.397, 0.336 and 0.267 respectively. The questions gathered for this study are representative of the SAT exams from 2008 to 2014, with the proportion of easy, medium, and hard questions in the bootstrapped SAT aligning with the average proportion in the complete SATs.

    Having predicted the difficulty of the questions, we use the predicted classes to identify the change in performance of the LLM agent over time. In Figure \ref{fig:difficulty_alignment}, we show the correct response rate of the LLM agent for easy, medium, and hard questions over time. First, we notice the LLM agent performs better on easy and medium questions compared to hard questions for the years 2008 to 2014. The agent finds those questions rated difficult for students to be challenging as well, highlighting the alignment of difficulty for students and LLM agent. The performance of the agent on easy, medium and hard questions in the predicted classes improves over time. This suggests, secondly, the LLM agent is able to answer questions predicted to be hard more accurately over time. The results are consistent with the decline in SAT math rigor over time.


\end{appendix}

\pagebreak

\section*{Tables}


\begin{table}[htbp]
    \centering
    \caption{Yearly SAT Score Changes}
    \label{tab:yearly_sat_decline}
    \begin{tabular}{lcccc}
        \toprule
        \toprule
        Year & \multicolumn{2}{c}{State} & \multicolumn{2}{c}{Massachusetts}                      \\
        \cmidrule(lr){2-3} \cmidrule(lr){4-5}
             & Estimate                  & S.E.                              & Estimate & S.E.    \\
        \midrule
        2009 & -1.831                    & (5.703)                           & 1.052    & (3.483) \\
        2010 & -32.471                   & (5.703)                           & -27.776  & (3.483) \\
        2011 & -39.792                   & (5.703)                           & -34.448  & (3.483) \\
        2012 & -13.357                   & (5.703)                           & -6.112   & (3.483) \\
        2013 & -25.980                   & (5.703)                           & -20.404  & (3.483) \\
        2014 & -23.271                   & (5.703)                           & -17.173  & (3.483) \\
        2015 & -17.063                   & (5.703)                           & -10.441  & (3.483) \\
        2016 & -18.784                   & (5.703)                           & -13.114  & (3.483) \\
        2017 & -33.416                   & (5.703)                           & -15.897  & (3.483) \\
        2018 & -31.643                   & (5.703)                           & -11.861  & (3.483) \\
        2019 & -46.522                   & (5.703)                           & -26.236  & (3.483) \\
        2020 & -83.749                   & (5.703)                           & -54.599  & (3.483) \\
        2021 & -55.439                   & (5.703)                           & -29.545  & (3.483) \\
        2022 & -77.737                   & (5.703)                           & -51.429  & (3.483) \\
        2023 & -99.471                   & (5.703)                           & -66.000  & (3.483) \\
        \bottomrule
        \bottomrule
    \end{tabular}
    \parbox{\textwidth}{ 
        \footnotesize
        Note: This table reports the decline in average SAT scores after considering transformed control for national data aggregated using unweighted \textit{State} SAT averages and \textit{Massachusetts} school district (Mass) level SAT. The estimates and standard errors (S.E.) are provided for each year from 2009 to 2023, with each year's comparison relative to 2008. The values represent the change in average SAT scores compared to the average performance of LLM agent, with standard errors indicating the variability of these estimates. The estimates for the national using the aggregated state and Massachusetts districts are based on unweighted SAT means each year.
    }
\end{table}

\begin{table}[htbp]
    \centering
    \caption{Yearly National, State, and Massachusetts SAT Score Changes}
    \label{tab:bayes_sat_decline}
    \begin{tabular}{lcccccc}
        \toprule
        \toprule
        Year & \multicolumn{2}{c}{National} & \multicolumn{2}{c}{State} & \multicolumn{2}{c}{Massachusetts}                                 \\
        \cmidrule(lr){2-3} \cmidrule(lr){4-5} \cmidrule(lr){6-7}
             & Estimate                     & S.E.                      & Estimate                          & S.E.     & Estimate & S.E.    \\
        \midrule
        2009 & -1.384                       & (4.757)                   & -1.871                            & (9.508)  & 1.012    & (6.511) \\
        2010 & -31.965                      & (4.22)                    & -32.53                            & (9.102)  & -27.761  & (6.128) \\
        2011 & -38.334                      & (4.837)                   & -39.806                           & (9.521)  & -34.413  & (6.498) \\
        2012 & -12.199                      & (5.111)                   & -13.396                           & (9.829)  & -6.066   & (6.686) \\
        2013 & -23.971                      & (4.913)                   & -26.008                           & (9.945)  & -20.34   & (6.636) \\
        2014 & -23.788                      & (4.541)                   & -23.418                           & (10.08)  & -17.163  & (6.358) \\
        2015 & -18.159                      & (5.195)                   & -17.198                           & (10.173) & -10.405  & (6.806) \\
        2016 & -24.986                      & (4.344)                   & -18.818                           & (10.122) & -13.154  & (6.183) \\
        2017 & -40.189                      & (4.937)                   & -33.352                           & (9.99)   & -15.89   & (6.499) \\
        2018 & -34.761                      & (4.346)                   & -31.702                           & (10.057) & -11.919  & (6.078) \\
        2019 & -47.576                      & (5.371)                   & -46.598                           & (10.857) & -26.192  & (6.88)  \\
        2020 & -83.177                      & (4.908)                   & -83.768                           & (10.201) & -54.554  & (6.604) \\
        2021 & -60.391                      & (4.23)                    & -55.552                           & (9.858)  & -29.56   & (6.145) \\
        2022 & -80.59                       & (4.336)                   & -77.797                           & (10.025) & -51.445  & (6.318) \\
        2023 & -106.993                     & (4.961)                   & -99.594                           & (10.834) & -65.936  & (6.75)  \\
        \bottomrule
        \bottomrule
    \end{tabular}
    \parbox{\textwidth}{ 
        \footnotesize
        This table reports the difference in average SAT scores after considering transformed control for data at different geographical levels using the Bayesian approach. The estimates and standard errors (S.E.) are provided for each year from 2009 to 2023 for the \textit{National} using the population parameters from the SAT reports, \textit{State} using national data aggregated from unweighted state SAT averages, and \textit{Massachusetts} using school district-wide SAT averages, with each year's comparison relative to 2008. The values represent the change in average SAT scores compared to the average performance of LLM agent, with standard errors indicating the variability of these estimates. Standard errors for national level comparison were produced using standard deviation and number of test-takers information provided in College Board's yearly SAT reports.
    }
\end{table}

\pagebreak


\renewcommand{\thetable}{A.\arabic{table}}
\setcounter{table}{0}

\begin{longtable}{cccccc}
    \caption{Concordance Tables for raw-to-scaled and pre-to-post score conversions}                                                                          \\
    \toprule
    \multicolumn{2}{c}{Post} & \multicolumn{2}{c}{Pre} & {Pre}                  & {Post}                                                                      \\
    \cmidrule(lr){1-2} \cmidrule(lr){3-4} \cmidrule(lr){5-5} \cmidrule(lr){6-6}
    {Raw Score}              & {Scaled Score}          & {Raw Score}            & {Scaled Score}          & {Scaled Score: Old}     & {Scaled Score: New}     \\
    \midrule
    \endfirsthead

    \multicolumn{6}{c}%
    {{\bfseries \tablename\ \thetable{} -- continued from previous page}}                                                                                     \\
    \toprule
    \multicolumn{2}{c}{Post} & \multicolumn{2}{c}{Pre} & {Pre}                  & {Post}                                                                      \\
    \cmidrule(lr){1-2} \cmidrule(lr){3-4} \cmidrule(lr){5-5} \cmidrule(lr){6-6}
    {Raw Score}              & {Scaled Score}          & {Raw Score}            & {Scaled Score}          & {Scaled Score: Old}     & {Scaled Score: New}     \\
    \midrule
    \endhead

    \bottomrule
    \endfoot

    \multicolumn{1}{c}{58}   & \multicolumn{1}{c}{800} & \multicolumn{1}{c}{54} & \multicolumn{1}{c}{800} & \multicolumn{1}{c}{200} & \multicolumn{1}{c}{200} \\
    \multicolumn{1}{c}{57}   & \multicolumn{1}{c}{790} & \multicolumn{1}{c}{53} & \multicolumn{1}{c}{790} & \multicolumn{1}{c}{210} & \multicolumn{1}{c}{220} \\
    \multicolumn{1}{c}{56}   & \multicolumn{1}{c}{780} & \multicolumn{1}{c}{52} & \multicolumn{1}{c}{760} & \multicolumn{1}{c}{220} & \multicolumn{1}{c}{230} \\
    \multicolumn{1}{c}{55}   & \multicolumn{1}{c}{760} & \multicolumn{1}{c}{51} & \multicolumn{1}{c}{740} & \multicolumn{1}{c}{230} & \multicolumn{1}{c}{250} \\
    \multicolumn{1}{c}{54}   & \multicolumn{1}{c}{750} & \multicolumn{1}{c}{50} & \multicolumn{1}{c}{720} & \multicolumn{1}{c}{240} & \multicolumn{1}{c}{260} \\
    \multicolumn{1}{c}{53}   & \multicolumn{1}{c}{740} & \multicolumn{1}{c}{49} & \multicolumn{1}{c}{710} & \multicolumn{1}{c}{250} & \multicolumn{1}{c}{280} \\
    \multicolumn{1}{c}{52}   & \multicolumn{1}{c}{730} & \multicolumn{1}{c}{48} & \multicolumn{1}{c}{700} & \multicolumn{1}{c}{260} & \multicolumn{1}{c}{300} \\
    \multicolumn{1}{c}{51}   & \multicolumn{1}{c}{710} & \multicolumn{1}{c}{47} & \multicolumn{1}{c}{690} & \multicolumn{1}{c}{270} & \multicolumn{1}{c}{310} \\
    \multicolumn{1}{c}{50}   & \multicolumn{1}{c}{700} & \multicolumn{1}{c}{46} & \multicolumn{1}{c}{680} & \multicolumn{1}{c}{280} & \multicolumn{1}{c}{330} \\
    \multicolumn{1}{c}{49}   & \multicolumn{1}{c}{690} & \multicolumn{1}{c}{45} & \multicolumn{1}{c}{670} & \multicolumn{1}{c}{290} & \multicolumn{1}{c}{340} \\
    \multicolumn{1}{c}{48}   & \multicolumn{1}{c}{680} & \multicolumn{1}{c}{44} & \multicolumn{1}{c}{660} & \multicolumn{1}{c}{300} & \multicolumn{1}{c}{350} \\
    \multicolumn{1}{c}{47}   & \multicolumn{1}{c}{670} & \multicolumn{1}{c}{43} & \multicolumn{1}{c}{650} & \multicolumn{1}{c}{310} & \multicolumn{1}{c}{360} \\
    \multicolumn{1}{c}{46}   & \multicolumn{1}{c}{670} & \multicolumn{1}{c}{42} & \multicolumn{1}{c}{640} & \multicolumn{1}{c}{320} & \multicolumn{1}{c}{360} \\
    \multicolumn{1}{c}{45}   & \multicolumn{1}{c}{660} & \multicolumn{1}{c}{41} & \multicolumn{1}{c}{640} & \multicolumn{1}{c}{330} & \multicolumn{1}{c}{370} \\
    \multicolumn{1}{c}{44}   & \multicolumn{1}{c}{650} & \multicolumn{1}{c}{40} & \multicolumn{1}{c}{630} & \multicolumn{1}{c}{340} & \multicolumn{1}{c}{380} \\
    \multicolumn{1}{c}{43}   & \multicolumn{1}{c}{640} & \multicolumn{1}{c}{39} & \multicolumn{1}{c}{620} & \multicolumn{1}{c}{350} & \multicolumn{1}{c}{390} \\
    \multicolumn{1}{c}{42}   & \multicolumn{1}{c}{630} & \multicolumn{1}{c}{38} & \multicolumn{1}{c}{610} & \multicolumn{1}{c}{360} & \multicolumn{1}{c}{400} \\
    \multicolumn{1}{c}{41}   & \multicolumn{1}{c}{620} & \multicolumn{1}{c}{37} & \multicolumn{1}{c}{600} & \multicolumn{1}{c}{370} & \multicolumn{1}{c}{410} \\
    \multicolumn{1}{c}{40}   & \multicolumn{1}{c}{610} & \multicolumn{1}{c}{36} & \multicolumn{1}{c}{590} & \multicolumn{1}{c}{380} & \multicolumn{1}{c}{420} \\
    \multicolumn{1}{c}{39}   & \multicolumn{1}{c}{600} & \multicolumn{1}{c}{35} & \multicolumn{1}{c}{590} & \multicolumn{1}{c}{390} & \multicolumn{1}{c}{430} \\
    \multicolumn{1}{c}{38}   & \multicolumn{1}{c}{600} & \multicolumn{1}{c}{34} & \multicolumn{1}{c}{580} & \multicolumn{1}{c}{400} & \multicolumn{1}{c}{440} \\
    \multicolumn{1}{c}{37}   & \multicolumn{1}{c}{590} & \multicolumn{1}{c}{33} & \multicolumn{1}{c}{570} & \multicolumn{1}{c}{410} & \multicolumn{1}{c}{450} \\
    \multicolumn{1}{c}{36}   & \multicolumn{1}{c}{580} & \multicolumn{1}{c}{32} & \multicolumn{1}{c}{560} & \multicolumn{1}{c}{420} & \multicolumn{1}{c}{460} \\
    \multicolumn{1}{c}{35}   & \multicolumn{1}{c}{570} & \multicolumn{1}{c}{31} & \multicolumn{1}{c}{550} & \multicolumn{1}{c}{430} & \multicolumn{1}{c}{470} \\
    \multicolumn{1}{c}{34}   & \multicolumn{1}{c}{560} & \multicolumn{1}{c}{30} & \multicolumn{1}{c}{540} & \multicolumn{1}{c}{440} & \multicolumn{1}{c}{480} \\
    \multicolumn{1}{c}{33}   & \multicolumn{1}{c}{560} & \multicolumn{1}{c}{29} & \multicolumn{1}{c}{540} & \multicolumn{1}{c}{450} & \multicolumn{1}{c}{490} \\
    \multicolumn{1}{c}{32}   & \multicolumn{1}{c}{550} & \multicolumn{1}{c}{28} & \multicolumn{1}{c}{530} & \multicolumn{1}{c}{460} & \multicolumn{1}{c}{500} \\
    \multicolumn{1}{c}{31}   & \multicolumn{1}{c}{540} & \multicolumn{1}{c}{27} & \multicolumn{1}{c}{520} & \multicolumn{1}{c}{470} & \multicolumn{1}{c}{510} \\
    \multicolumn{1}{c}{30}   & \multicolumn{1}{c}{530} & \multicolumn{1}{c}{26} & \multicolumn{1}{c}{510} & \multicolumn{1}{c}{480} & \multicolumn{1}{c}{510} \\
    \multicolumn{1}{c}{29}   & \multicolumn{1}{c}{520} & \multicolumn{1}{c}{25} & \multicolumn{1}{c}{500} & \multicolumn{1}{c}{490} & \multicolumn{1}{c}{520} \\
    \multicolumn{1}{c}{28}   & \multicolumn{1}{c}{520} & \multicolumn{1}{c}{24} & \multicolumn{1}{c}{490} & \multicolumn{1}{c}{500} & \multicolumn{1}{c}{530} \\
    \multicolumn{1}{c}{27}   & \multicolumn{1}{c}{510} & \multicolumn{1}{c}{23} & \multicolumn{1}{c}{480} & \multicolumn{1}{c}{510} & \multicolumn{1}{c}{540} \\
    \multicolumn{1}{c}{26}   & \multicolumn{1}{c}{500} & \multicolumn{1}{c}{22} & \multicolumn{1}{c}{480} & \multicolumn{1}{c}{520} & \multicolumn{1}{c}{550} \\
    \multicolumn{1}{c}{25}   & \multicolumn{1}{c}{490} & \multicolumn{1}{c}{21} & \multicolumn{1}{c}{470} & \multicolumn{1}{c}{530} & \multicolumn{1}{c}{560} \\
    \multicolumn{1}{c}{24}   & \multicolumn{1}{c}{480} & \multicolumn{1}{c}{20} & \multicolumn{1}{c}{460} & \multicolumn{1}{c}{540} & \multicolumn{1}{c}{570} \\
    \multicolumn{1}{c}{23}   & \multicolumn{1}{c}{480} & \multicolumn{1}{c}{19} & \multicolumn{1}{c}{450} & \multicolumn{1}{c}{550} & \multicolumn{1}{c}{570} \\
    \multicolumn{1}{c}{22}   & \multicolumn{1}{c}{470} & \multicolumn{1}{c}{18} & \multicolumn{1}{c}{440} & \multicolumn{1}{c}{560} & \multicolumn{1}{c}{580} \\
    \multicolumn{1}{c}{21}   & \multicolumn{1}{c}{460} & \multicolumn{1}{c}{17} & \multicolumn{1}{c}{430} & \multicolumn{1}{c}{570} & \multicolumn{1}{c}{590} \\
    \multicolumn{1}{c}{20}   & \multicolumn{1}{c}{450} & \multicolumn{1}{c}{16} & \multicolumn{1}{c}{420} & \multicolumn{1}{c}{580} & \multicolumn{1}{c}{600} \\
    \multicolumn{1}{c}{19}   & \multicolumn{1}{c}{440} & \multicolumn{1}{c}{15} & \multicolumn{1}{c}{420} & \multicolumn{1}{c}{590} & \multicolumn{1}{c}{610} \\
    \multicolumn{1}{c}{18}   & \multicolumn{1}{c}{430} & \multicolumn{1}{c}{14} & \multicolumn{1}{c}{410} & \multicolumn{1}{c}{600} & \multicolumn{1}{c}{620} \\
    \multicolumn{1}{c}{17}   & \multicolumn{1}{c}{420} & \multicolumn{1}{c}{13} & \multicolumn{1}{c}{400} & \multicolumn{1}{c}{610} & \multicolumn{1}{c}{630} \\
    \multicolumn{1}{c}{16}   & \multicolumn{1}{c}{410} & \multicolumn{1}{c}{12} & \multicolumn{1}{c}{390} & \multicolumn{1}{c}{620} & \multicolumn{1}{c}{640} \\
    \multicolumn{1}{c}{15}   & \multicolumn{1}{c}{390} & \multicolumn{1}{c}{11} & \multicolumn{1}{c}{380} & \multicolumn{1}{c}{630} & \multicolumn{1}{c}{650} \\
    \multicolumn{1}{c}{14}   & \multicolumn{1}{c}{380} & \multicolumn{1}{c}{10} & \multicolumn{1}{c}{370} & \multicolumn{1}{c}{640} & \multicolumn{1}{c}{660} \\
    \multicolumn{1}{c}{13}   & \multicolumn{1}{c}{370} & \multicolumn{1}{c}{9}  & \multicolumn{1}{c}{360} & \multicolumn{1}{c}{650} & \multicolumn{1}{c}{670} \\
    \multicolumn{1}{c}{12}   & \multicolumn{1}{c}{360} & \multicolumn{1}{c}{8}  & \multicolumn{1}{c}{350} & \multicolumn{1}{c}{660} & \multicolumn{1}{c}{690} \\
    \multicolumn{1}{c}{11}   & \multicolumn{1}{c}{340} & \multicolumn{1}{c}{7}  & \multicolumn{1}{c}{330} & \multicolumn{1}{c}{670} & \multicolumn{1}{c}{700} \\
    \multicolumn{1}{c}{10}   & \multicolumn{1}{c}{330} & \multicolumn{1}{c}{6}  & \multicolumn{1}{c}{320} & \multicolumn{1}{c}{680} & \multicolumn{1}{c}{710} \\
    \multicolumn{1}{c}{9}    & \multicolumn{1}{c}{320} & \multicolumn{1}{c}{5}  & \multicolumn{1}{c}{310} & \multicolumn{1}{c}{690} & \multicolumn{1}{c}{720} \\
    \multicolumn{1}{c}{8}    & \multicolumn{1}{c}{310} & \multicolumn{1}{c}{4}  & \multicolumn{1}{c}{290} & \multicolumn{1}{c}{700} & \multicolumn{1}{c}{730} \\
    \multicolumn{1}{c}{7}    & \multicolumn{1}{c}{290} & \multicolumn{1}{c}{3}  & \multicolumn{1}{c}{280} & \multicolumn{1}{c}{710} & \multicolumn{1}{c}{740} \\
    \multicolumn{1}{c}{6}    & \multicolumn{1}{c}{280} & \multicolumn{1}{c}{2}  & \multicolumn{1}{c}{260} & \multicolumn{1}{c}{720} & \multicolumn{1}{c}{750} \\
    \multicolumn{1}{c}{5}    & \multicolumn{1}{c}{260} & \multicolumn{1}{c}{1}  & \multicolumn{1}{c}{240} & \multicolumn{1}{c}{730} & \multicolumn{1}{c}{760} \\
    \multicolumn{1}{c}{4}    & \multicolumn{1}{c}{240} & \multicolumn{1}{c}{0}  & \multicolumn{1}{c}{220} & \multicolumn{1}{c}{740} & \multicolumn{1}{c}{760} \\
    \multicolumn{1}{c}{3}    & \multicolumn{1}{c}{230} & \multicolumn{1}{c}{-1} & \multicolumn{1}{c}{200} & \multicolumn{1}{c}{750} & \multicolumn{1}{c}{770} \\
    \multicolumn{1}{c}{2}    & \multicolumn{1}{c}{210} & \multicolumn{1}{c}{-2} & \multicolumn{1}{c}{200} & \multicolumn{1}{c}{760} & \multicolumn{1}{c}{780} \\
    \multicolumn{1}{c}{1}    & \multicolumn{1}{c}{200} & \multicolumn{1}{c}{}   & \multicolumn{1}{c}{}    & \multicolumn{1}{c}{770} & \multicolumn{1}{c}{780} \\
    \multicolumn{1}{c}{0}    & \multicolumn{1}{c}{200} & \multicolumn{1}{c}{}   & \multicolumn{1}{c}{}    & \multicolumn{1}{c}{780} & \multicolumn{1}{c}{790} \\
    \multicolumn{1}{c}{}     & \multicolumn{1}{c}{}    & \multicolumn{1}{c}{}   & \multicolumn{1}{c}{}    & \multicolumn{1}{c}{790} & \multicolumn{1}{c}{800} \\
    \multicolumn{1}{c}{}     & \multicolumn{1}{c}{}    & \multicolumn{1}{c}{}   & \multicolumn{1}{c}{}    & \multicolumn{1}{c}{800} & \multicolumn{1}{c}{800}
\end{longtable}

\renewcommand{\thetable}{D.\arabic{table}}
\setcounter{table}{0}

\begin{table}[htbp]
    \centering
    \caption{Yearly National, State, and Massachusetts SAT Score Changes using GPT4 April}
    \label{tab:bayes_sat_decline_gpt4}
    \begin{tabular}{lcccccc}
        \toprule
        \toprule
        Year & \multicolumn{2}{c}{National} & \multicolumn{2}{c}{State} & \multicolumn{2}{c}{Massachusetts}                                 \\
        \cmidrule(lr){2-3} \cmidrule(lr){4-5} \cmidrule(lr){6-7}
             & Estimate                     & S.E.                      & Estimate                          & S.E.     & Estimate & S.E.    \\
        \midrule
        2009 & -0.01                        & (4.506)                   & 0.011                             & (9.148)  & 2.997    & (6.298) \\
        2010 & -32.016                      & (4.009)                   & -31.97                            & (8.856)  & -27.987  & (5.931) \\
        2011 & -32.996                      & (4.402)                   & -34.035                           & (9.196)  & -28.986  & (6.151) \\
        2012 & -13.016                      & (4.701)                   & -14.049                           & (9.449)  & -6.997   & (6.353) \\
        2013 & -18.999                      & (4.705)                   & -21.04                            & (9.647)  & -14.985  & (6.38)  \\
        2014 & -41.004                      & (3.783)                   & -39.985                           & (9.492)  & -33.992  & (5.793) \\
        2015 & -25.986                      & (4.709)                   & -24.942                           & (9.833)  & -18.008  & (6.369) \\
        2016 & -22.004                      & (3.919)                   & -15.905                           & (9.777)  & -10.013  & (5.867) \\
        2017 & -38.0                        & (4.296)                   & -31.049                           & (9.755)  & -14.009  & (5.962) \\
        2018 & -22.009                      & (4.103)                   & -18.995                           & (9.849)  & 0.971    & (5.894) \\
        2019 & -35.971                      & (4.814)                   & -35.073                           & (10.392) & -13.983  & (6.394) \\
        2020 & -71.98                       & (4.713)                   & -72.03                            & (9.869)  & -43.032  & (6.404) \\
        2021 & -64.993                      & (4.0)                     & -60.039                           & (9.486)  & -33.978  & (5.944) \\
        2022 & -92.994                      & (3.849)                   & -89.983                           & (9.635)  & -63.969  & (5.895) \\
        2023 & -113.011                     & (4.572)                   & -105.008                          & (10.488) & -72.031  & (6.43)  \\
        \bottomrule
        \bottomrule
    \end{tabular}
    \parbox{\textwidth}{ 
        \footnotesize
        This table reports the difference in average SAT scores after considering transformed control using GPT4 April update for data at different geographical levels using the Bayesian approach. The estimates and standard errors (S.E.) are provided for each year from 2009 to 2023 for the \textit{National} using the population parameters from the SAT reports, \textit{State} using national data aggregated from unweighted state SAT averages, and \textit{Massachusetts} using school district-wide SAT averages, with each year's comparison relative to 2008. The values represent the change in average SAT scores compared to the average performance of LLM agent, with standard errors indicating the variability of these estimates. Standard errors for national level comparison were produced using standard deviation and number of test-takers information provided in College Board's yearly SAT reports.
    }    
\end{table}

\renewcommand{\thetable}{E.\arabic{table}}
\setcounter{table}{0}

\begin{table}[!htbp]
    \centering
    \caption{Yearly Performance of LLM Agent by Question Type}
    \label{tab:yearly_performance}
    \begin{tabular}{lcccccc}
        \toprule
        \toprule
        Year & \multicolumn{2}{c}{All Questions} & \multicolumn{2}{c}{Multiple Choice only} & \multicolumn{2}{c}{Answer Type only}                                  \\
        \cmidrule(lr){2-3} \cmidrule(lr){4-5} \cmidrule(lr){6-7}
             & Proportion                        & S.E.                                     & Proportion                           & S.E.    & Proportion & S.E.    \\
        \midrule
        2008 & 0.469                             & (0.008)                                  & 0.495                                & (0.009) & 0.378      & (0.018) \\
        2009 & 0.473                             & (0.007)                                  & 0.459                                & (0.008) & 0.523      & (0.013) \\
        2010 & 0.548                             & (0.006)                                  & 0.525                                & (0.007) & 0.630      & (0.013) \\
        2011 & 0.560                             & (0.008)                                  & 0.555                                & (0.009) & 0.578      & (0.011) \\
        2012 & 0.494                             & (0.008)                                  & 0.493                                & (0.009) & 0.500      & (0.014) \\
        2013 & 0.523                             & (0.008)                                  & 0.546                                & (0.009) & 0.445      & (0.013) \\
        2014 & 0.519                             & (0.006)                                  & 0.543                                & (0.008) & 0.435      & (0.013) \\
        2015 & 0.502                             & (0.008)                                  & 0.517                                & (0.010) & 0.448      & (0.015) \\
        2016 & 0.508                             & (0.005)                                  & 0.516                                & (0.006) & 0.478      & (0.007) \\
        2017 & 0.574                             & (0.007)                                  & 0.603                                & (0.008) & 0.474      & (0.015) \\
        2018 & 0.569                             & (0.005)                                  & 0.592                                & (0.006) & 0.492      & (0.009) \\
        2019 & 0.592                             & (0.008)                                  & 0.630                                & (0.008) & 0.462      & (0.018) \\
        2020 & 0.657                             & (0.007)                                  & 0.654                                & (0.008) & 0.669      & (0.015) \\
        2021 & 0.619                             & (0.004)                                  & 0.578                                & (0.006) & 0.762      & (0.005) \\
        2022 & 0.648                             & (0.005)                                  & 0.672                                & (0.006) & 0.566      & (0.011) \\
        2023 & 0.675                             & (0.007)                                  & 0.699                                & (0.007) & 0.592      & (0.018) \\
        \bottomrule
        \bottomrule
    \end{tabular}
    \parbox{\textwidth}{ 
        \footnotesize
        This table reports the average proportion of questions correctly answered by the LLM agent each year. The estimates and standard errors (S.E.) for the proportion of questions correctly answers from a complete exam including questions of all types, multiple choice type only and answer type only are provided for each year from 2009 to 2023. The values represent the ratio of correctly answered questions to the total number of questions in that category of question type.
    }        
\end{table}

\begin{table}[!htbp]
    \centering
    \caption{Asian, Black and White Student SAT Score Changes}
    \label{tab:race_ads}
    \begin{tabular}{lcccccc}
        \toprule
        \toprule
        Year & \multicolumn{2}{c}{Asian} & \multicolumn{2}{c}{Black} & \multicolumn{2}{c}{White}                                \\
        \cmidrule(lr){2-3} \cmidrule(lr){4-5} \cmidrule(lr){6-7}
             & Estimate                  & S.E.                      & Estimate                  & S.E.    & Estimate & S.E.    \\
        \midrule
        2009 & 3.595                     & (4.764)                   & -1.326                    & (4.74)  & -2.493   & (4.73)  \\
        2010 & -22.978                   & (4.255)                   & -29.926                   & (4.204) & -33.027  & (4.203) \\
        2011 & -24.386                   & (4.797)                   & -36.367                   & (4.818) & -39.425  & (4.819) \\
        2012 & 1.832                     & (5.099)                   & -9.17                     & (5.039) & -12.268  & (5.001) \\
        2013 & -7.993                    & (4.945)                   & -19.935                   & (4.881) & -26.079  & (4.899) \\
        2014 & -5.761                    & (4.599)                   & -18.729                   & (4.609) & -24.826  & (4.623) \\
        2015 & 1.808                     & (5.195)                   & -12.163                   & (5.177) & -17.219  & (5.179) \\
        2016 & 1.993                     & (4.331)                   & -18.94                    & (4.295) & -22.032  & (4.329) \\
        2017 & -19.174                   & (4.988)                   & -18.104                   & (4.88)  & -35.219  & (4.969) \\
        2018 & 5.208                     & (4.371)                   & -15.736                   & (4.318) & -29.852  & (4.351) \\
        2019 & -2.59                     & (5.358)                   & -31.52                    & (5.423) & -43.61   & (5.454) \\
        2020 & -38.206                   & (4.904)                   & -65.152                   & (4.859) & -80.246  & (4.873) \\
        2021 & -10.403                   & (4.171)                   & -44.337                   & (4.179) & -59.447  & (4.248) \\
        2022 & -32.568                   & (4.31)                    & -62.526                   & (4.347) & -79.617  & (4.302) \\
        2023 & -52.988                   & (4.924)                   & -86.969                   & (4.898) & -103.99  & (4.892) \\
        \bottomrule
        \bottomrule
    \end{tabular}
\end{table}

\begin{table}[!h]
    \centering
    \caption{Male and Female Student SAT Score Changes}
    \label{tab:gender_ads}
    \begin{tabular}{lcccc}
        \toprule
        \toprule
        Year & \multicolumn{2}{c}{Male} & \multicolumn{2}{c}{Female}                      \\
        \cmidrule(lr){2-3} \cmidrule(lr){4-5}
             & Estimate                 & S.E.                       & Estimate & S.E.    \\
        \midrule
        2009 & -0.385                   & (4.766)                    & -2.353   & (4.725) \\
        2010 & -30.991                  & (4.248)                    & -32.007  & (4.207) \\
        2011 & -39.401                  & (4.847)                    & -37.417  & (4.84)  \\
        2012 & -12.23                   & (5.051)                    & -12.199  & (5.055) \\
        2013 & -24.989                  & (4.923)                    & -23.996  & (4.891) \\
        2014 & -24.796                  & (4.688)                    & -22.831  & (4.581) \\
        2015 & -20.182                  & (5.176)                    & -18.197  & (5.152) \\
        2016 & -26.988                  & (4.35)                     & -23.984  & (4.292) \\
        2017 & -46.147                  & (4.863)                    & -36.227  & (4.925) \\
        2018 & -40.797                  & (4.357)                    & -28.773  & (4.299) \\
        2019 & -55.57                   & (5.359)                    & -41.587  & (5.356) \\
        2020 & -92.155                  & (4.91)                     & -75.156  & (4.913) \\
        2021 & -68.379                  & (4.234)                    & -54.405  & (4.151) \\
        2022 & -88.565                  & (4.349)                    & -74.586  & (4.293) \\
        2023 & -117.006                 & (4.897)                    & -100.039 & (4.876) \\
        \bottomrule
        \bottomrule
    \end{tabular}
\end{table}

\begin{table}[!h]
    \centering
    \caption{Yearly National and Massachusetts SAT Score Changes after removing post-COVID years}
    \label{tab:yearly_sat_decline_post_covid}
    \begin{tabular}{lcccc}
        \toprule
        \toprule
        Year & \multicolumn{2}{c}{National} & \multicolumn{2}{c}{Massachusetts}                      \\
        \cmidrule(lr){2-3} \cmidrule(lr){4-5}
             & Estimate                     & S.E.                              & Estimate & S.E.    \\
        \midrule
        2009 & -1.831                       & (5.495)                           & 1.052    & (3.208) \\
        2010 & -32.471                      & (5.495)                           & -27.776  & (3.208) \\
        2011 & -39.792                      & (5.495)                           & -34.448  & (3.208) \\
        2012 & -13.357                      & (5.495)                           & -6.112   & (3.208) \\
        2013 & -25.980                      & (5.495)                           & -20.404  & (3.208) \\
        2014 & -23.271                      & (5.495)                           & -17.173  & (3.208) \\
        2015 & -17.063                      & (5.495)                           & -10.441  & (3.208) \\
        2016 & -18.784                      & (5.495)                           & -13.114  & (3.208) \\
        2017 & -33.416                      & (5.495)                           & -15.897  & (3.208) \\
        2018 & -31.643                      & (5.495)                           & -11.861  & (3.208) \\
        2019 & -46.522                      & (5.495)                           & -26.236  & (3.208) \\
        2020 & -83.749                      & (5.495)                           & -54.599  & (3.208) \\
        \bottomrule
        \bottomrule
    \end{tabular}
    \parbox{\textwidth}{ 
        \footnotesize
        This table reports results similar to Table \ref{tab:yearly_sat_decline} after removing post-COVID years. The national sample represented by the unweighted state SAT data.
    }           
\end{table}

\renewcommand{\thetable}{F.\arabic{table}}
\setcounter{table}{0}

\begin{table}[ht]
    \centering
    \caption{SAT math scores for Lindner College of Business entrants as per the faculty}
    \begin{tabularx}{\textwidth}{X>{\centering\arraybackslash}X>{\centering\arraybackslash}X}
        \toprule
        \toprule
                                                                                                                                           & \textbf{2008} & \textbf{2023} \\
        \midrule
        What is the average score you believe a high-school graduate entering the Lindner College of Business in 2008 would have received? & 573.83        & \textit{N/A}  \\
        \addlinespace
        What is the average score you believe a high-school graduate entering the Lindner College of Business in 2023 would have received? & \textit{N/A}  & 540.71        \\
        \addlinespace
        Standard Deviation                                                                                                                 & 69.41         & 84.81         \\
        Sample Size                                                                                                                        & 42            & 42            \\
        \bottomrule
        \bottomrule
    \end{tabularx}
    \label{table:sat_scores_combined}
\end{table}

\renewcommand{\thetable}{G.\arabic{table}}
\setcounter{table}{0}

\begin{table}[ht]
    \centering
    \caption{Confusion Matrix: Classification Accuracy - Test Sample}
    \begin{tabular}{lcccc}
        \hline
        \hline
        \textbf{Predicted/Ground Truth} & \textbf{Easy} & \textbf{Medium} & \textbf{Hard} \\
        \hline
        \textbf{Easy}                   & 130           & 15              & 1             \\
        \textbf{Medium}                 & 17            & 61              & 23            \\
        \textbf{Hard}                   & 1             & 4               & 36            \\
        \hline
        \hline
    \end{tabular}
    \label{table:confusion_matrix}
\end{table}

\pagebreak

\clearpage

\pagebreak

\section*{Figures}


\begin{figure}[ht]
    \centering
    \includegraphics[width=0.6\textwidth,keepaspectratio]{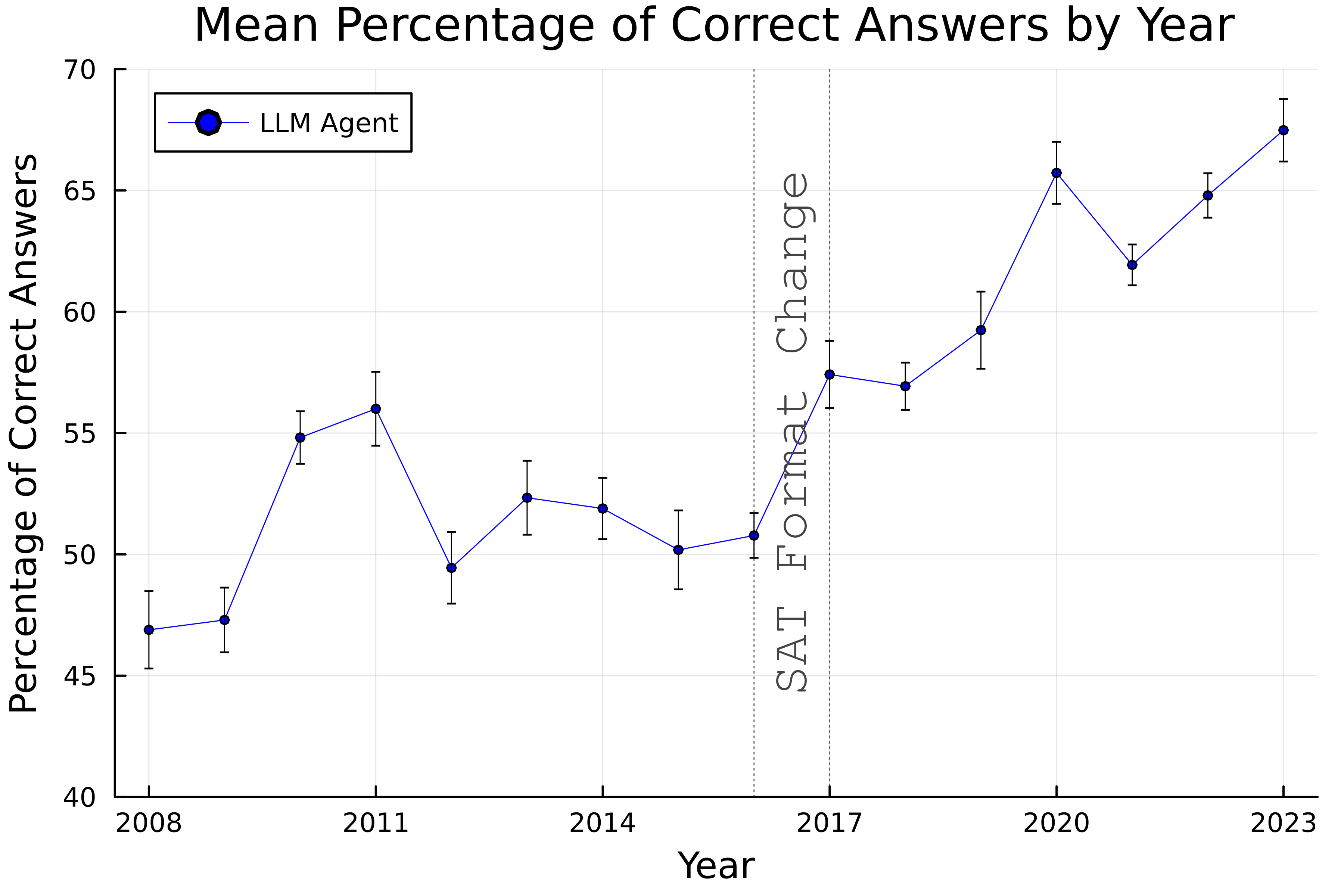}
    \caption{\% of questions correctly answered by LLM agent in SATs from each year}
    \label{fig:pct_gpt_score}
\end{figure}

\renewcommand{\thefigure}{B.\arabic{figure}}
\setcounter{figure}{0}

\begin{figure}[ht]
    \centering
    \includegraphics[width=0.6\textwidth,keepaspectratio]{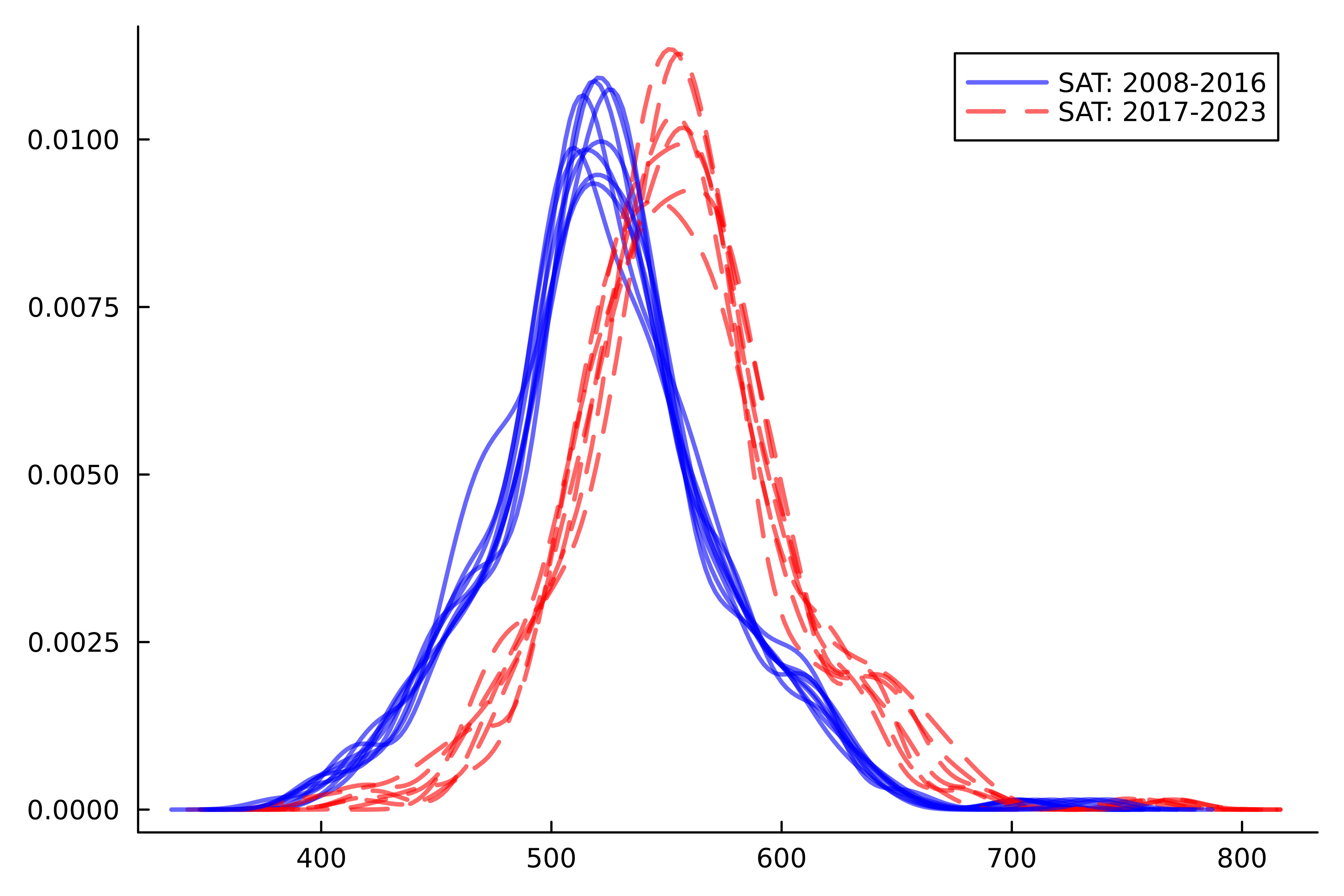}
    \caption{Before Conversion: SAT Score Distribution for School Districts in Massachusetts}
    \label{fig:avg_stud_scores_mass}
\end{figure}
\begin{figure}[ht]
    \centering
    \includegraphics[width=0.6\textwidth,keepaspectratio]{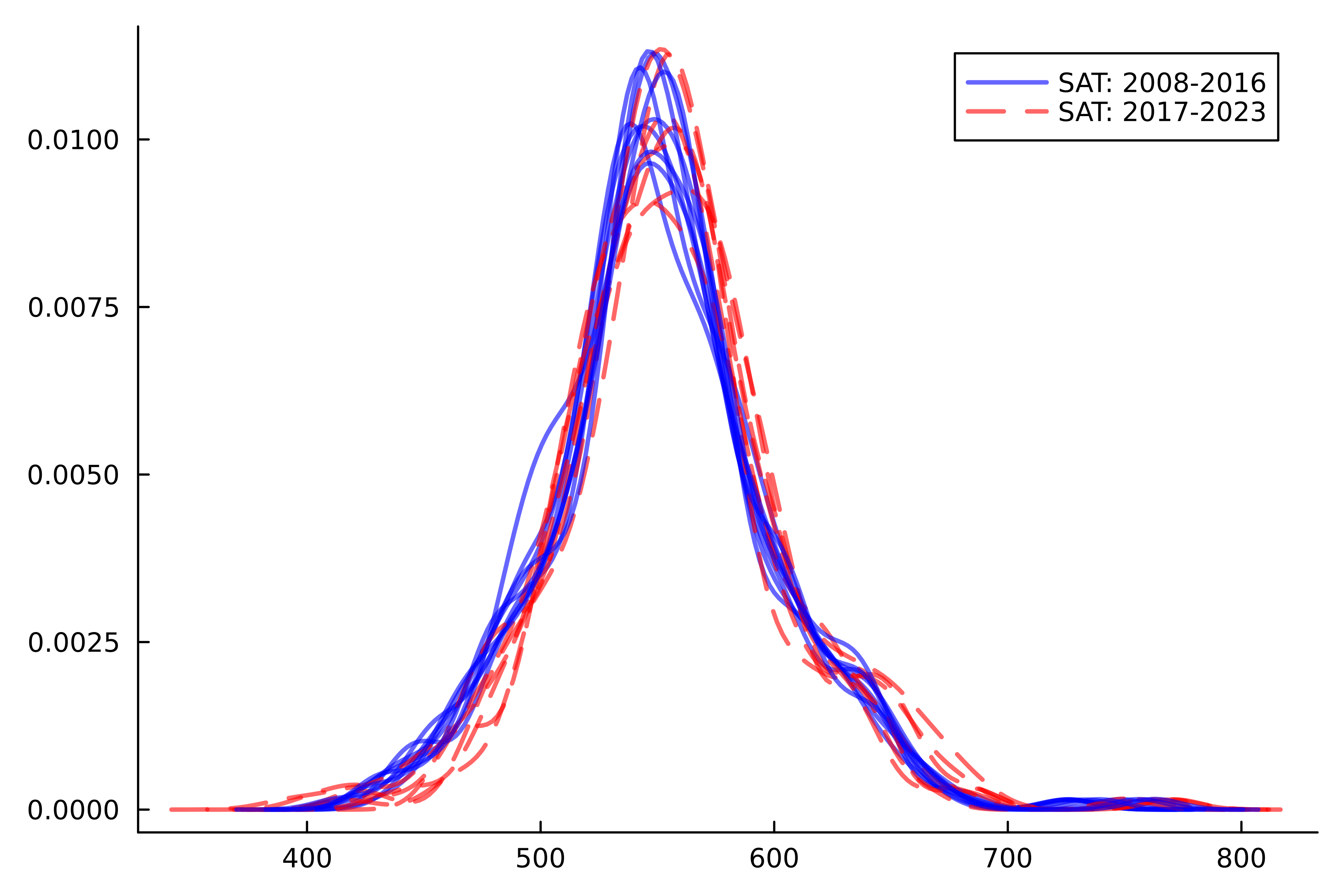}
    \caption{After Conversion: SAT Score Distribution for School Districts in Massachusetts}
    \label{fig:avg_stud_scores_mass_conv}
\end{figure}

\renewcommand{\thefigure}{D.\arabic{figure}}
\setcounter{figure}{0}

\begin{figure}[htbp]
    \centering
    \includegraphics[width=\textwidth,keepaspectratio]{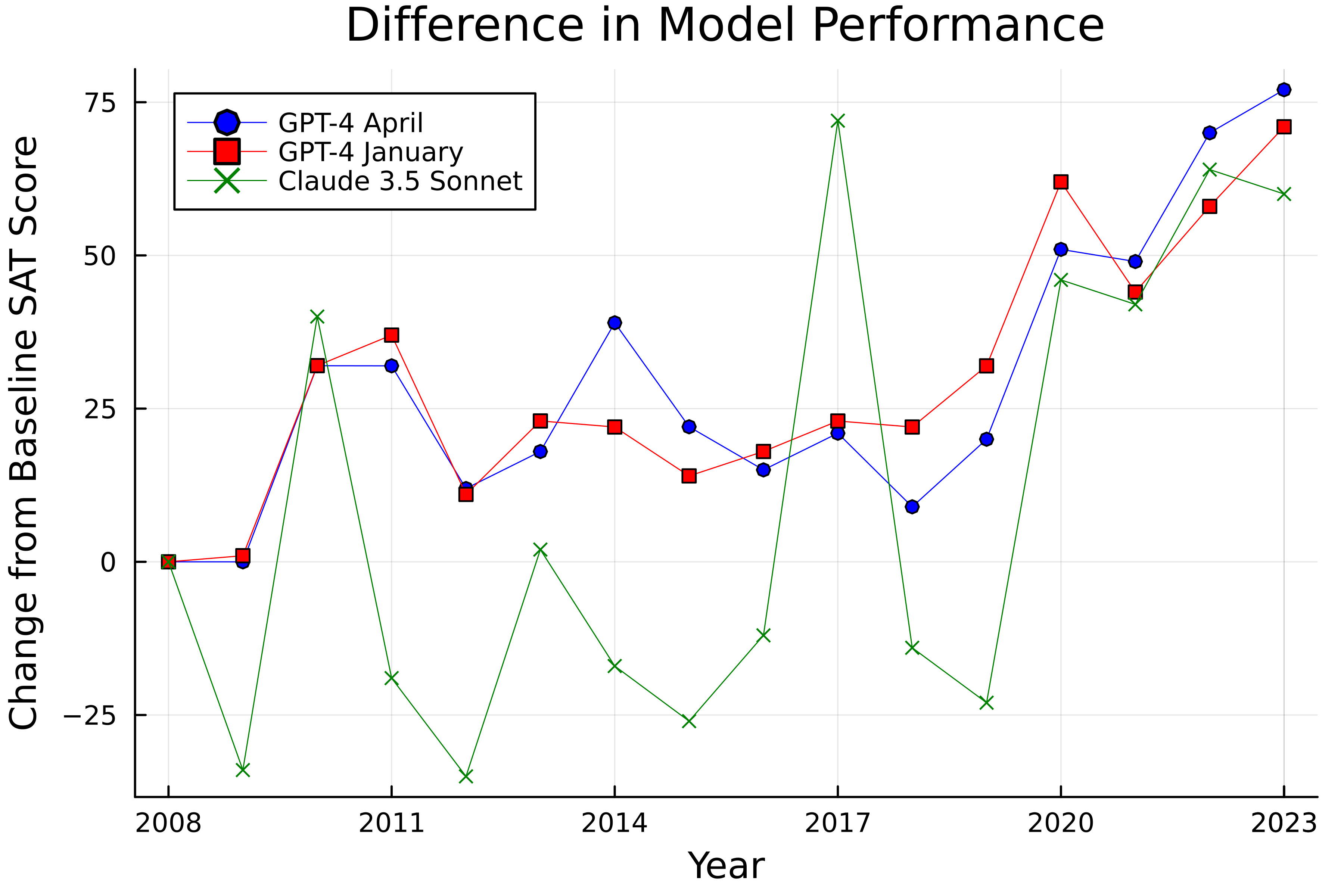}
    \caption{Difference in Performance of GPT-4 January, GPT-4 April and Claude 3.5 Sonnet}
    \label{fig:gpt4_april_jan_claude_diff}
\end{figure}

\renewcommand{\thefigure}{E.\arabic{figure}}
\setcounter{figure}{0}

\begin{figure}[ht]
    \centering
    \includegraphics[width=\textwidth,keepaspectratio]{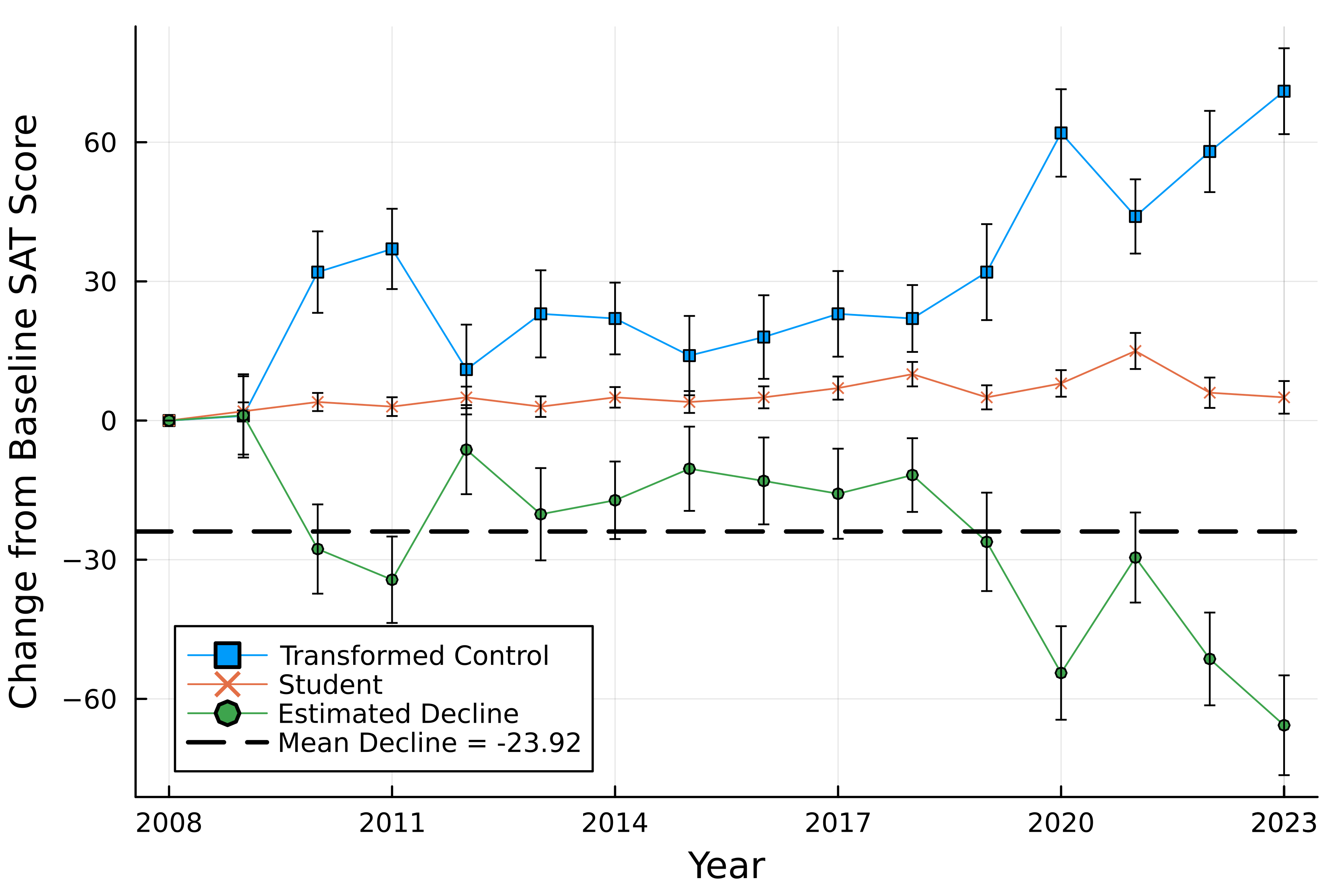}
    \caption{Bayesian Estimated $ADS(\hat{t},t)$ using Massachusetts district SAT data}
    \label{fig:bayes_mass_nopool}
\end{figure}

\begin{figure}[htbp]
    \centering
    \includegraphics[width=\textwidth,keepaspectratio]{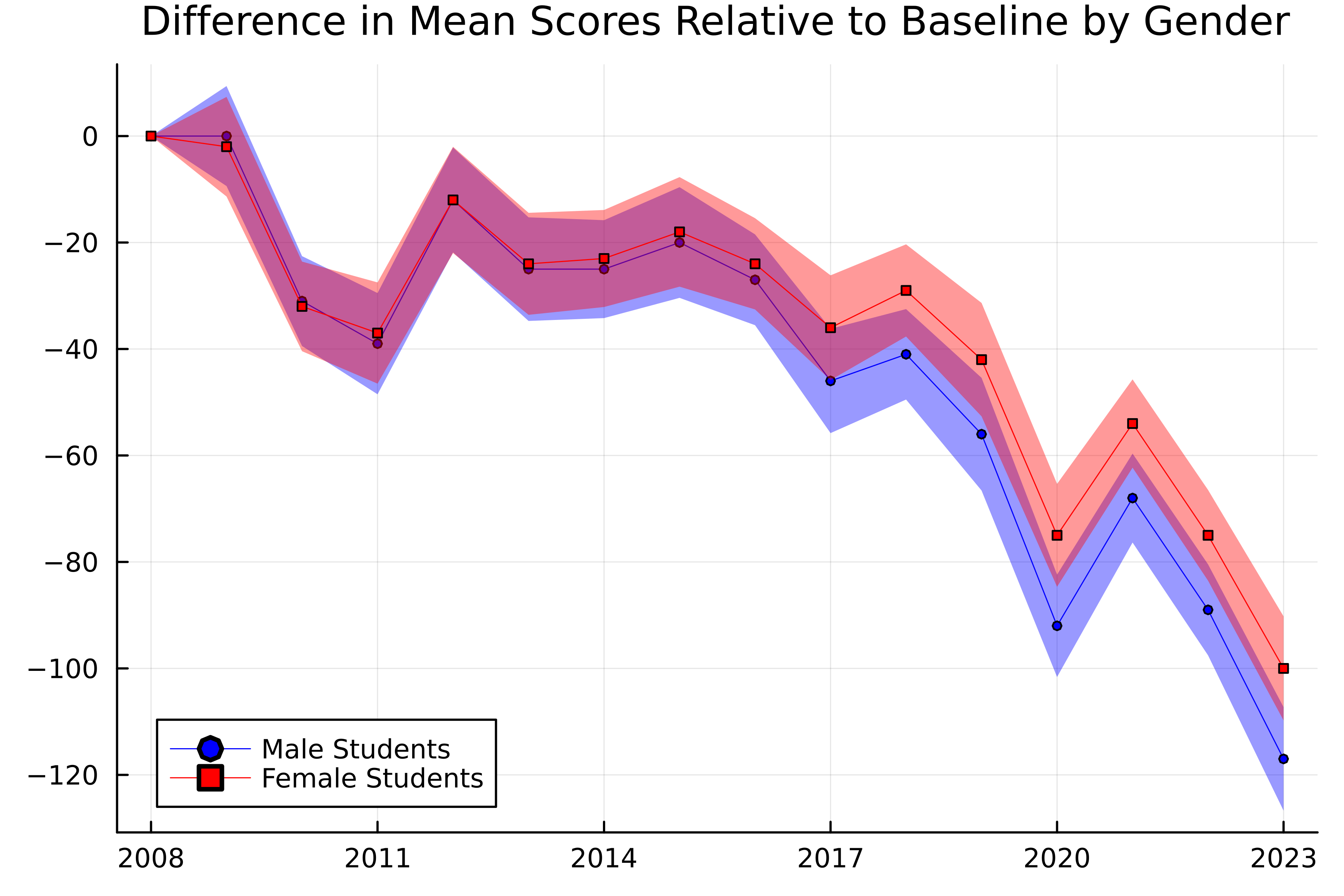}
    \caption{Estimated ADS for Male and Female}
    \label{fig:gender_ads}
\end{figure}

\renewcommand{\thefigure}{F.\arabic{figure}}
\setcounter{figure}{0}

\begin{figure}[!htbp]
    \centering
    \includegraphics[width=0.9\textwidth,keepaspectratio]{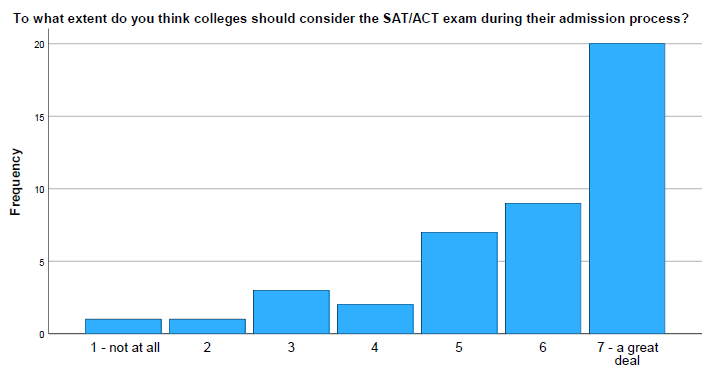}
    \caption{Faculty Response for Question 1}
    \label{fig:survey1_faculty}
\end{figure}
\begin{figure}[!htbp]
    \centering
    \includegraphics[width=0.9\textwidth,keepaspectratio]{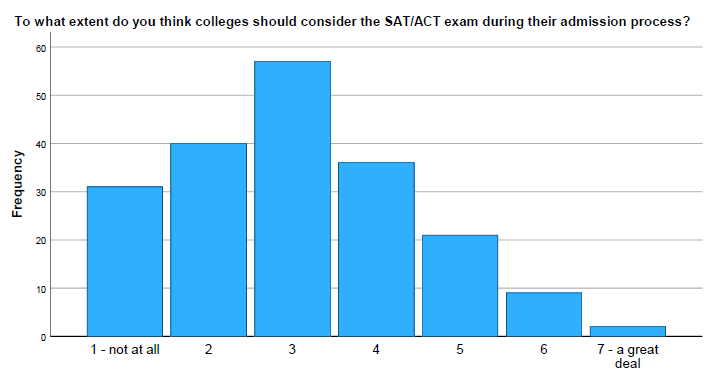}
    \caption{Student Response for Question 1}
    \label{fig:survey1_student}
\end{figure}

\begin{figure}[!htbp]
    \centering
    \includegraphics[width=0.9\textwidth,keepaspectratio]{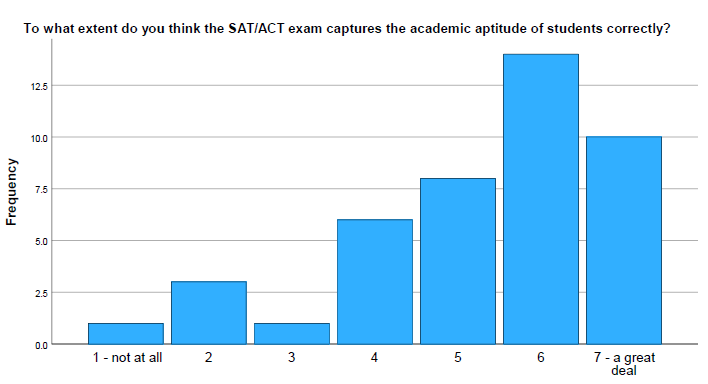}
    \caption{Faculty Response for Question 2}
    \label{fig:survey2_faculty}
\end{figure}
\begin{figure}[!htbp]
    \centering
    \includegraphics[width=0.9\textwidth,keepaspectratio]{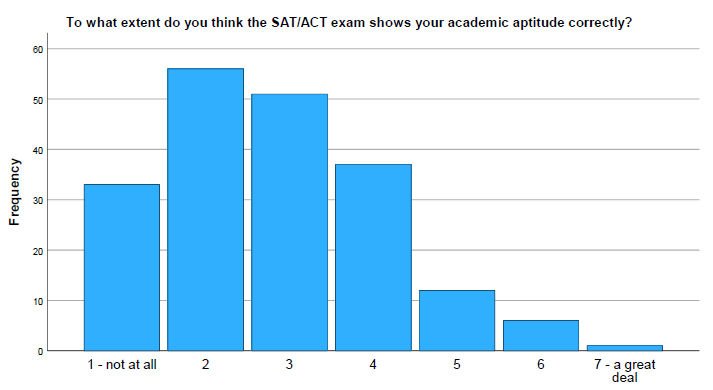}
    \caption{Student Response for Question 2}
    \label{fig:survey2_student}
\end{figure}

\begin{figure}[!htbp]
    \centering
    \includegraphics[width=0.9\textwidth,keepaspectratio]{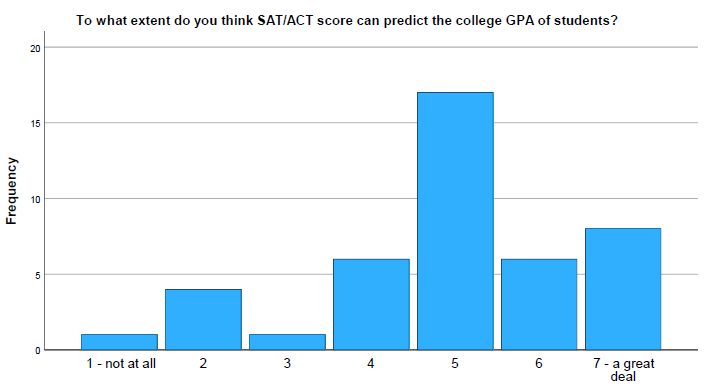}
    \caption{Faculty Response for Question 3}
    \label{fig:survey3_faculty}
\end{figure}
\begin{figure}[!htbp]
    \centering
    \includegraphics[width=0.9\textwidth,keepaspectratio]{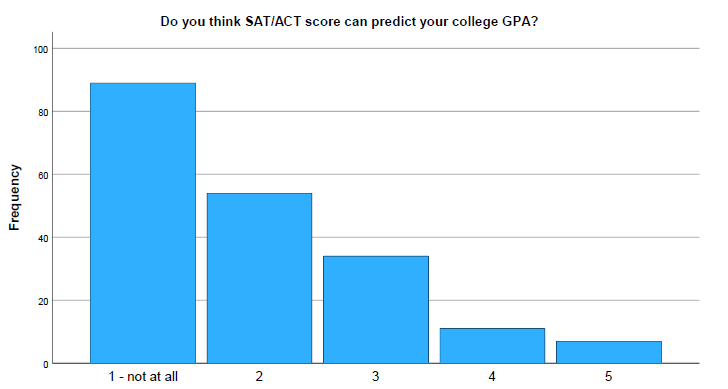}
    \caption{Student Response for Question 3}
    \label{fig:survey3_student}
\end{figure}

\begin{figure}[!htbp]
    \centering
    \includegraphics[width=0.9\textwidth,keepaspectratio]{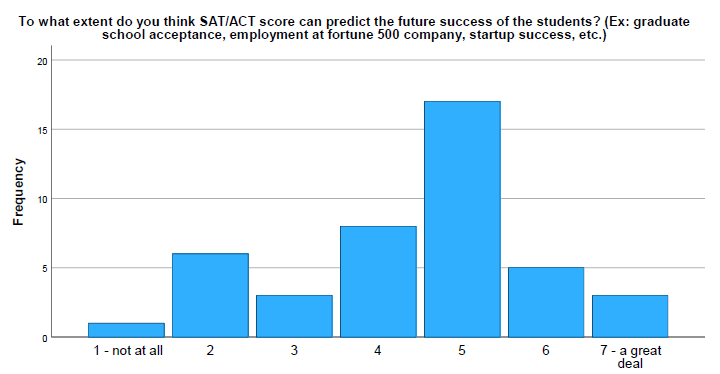}
    \caption{Faculty Response for Question 4}
    \label{fig:survey4_faculty}
\end{figure}
\begin{figure}[!htbp]
    \centering
    \includegraphics[width=0.9\textwidth,keepaspectratio]{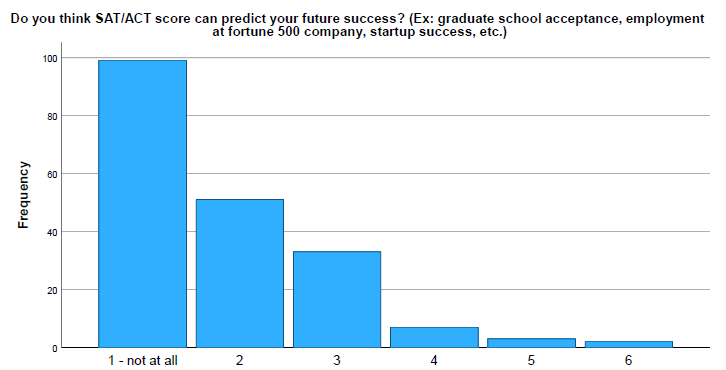}
    \caption{Student Response for Question 4}
    \label{fig:survey4_student}
\end{figure}

\begin{figure}[!htbp]
    \centering
    \includegraphics[width=0.9\textwidth,keepaspectratio]{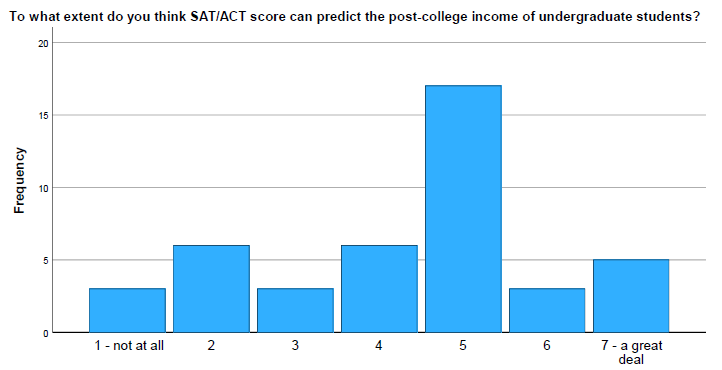}
    \caption{Faculty Response for Question 5}
    \label{fig:survey5_faculty}
\end{figure}
\begin{figure}[!htbp]
    \centering
    \includegraphics[width=0.9\textwidth,keepaspectratio]{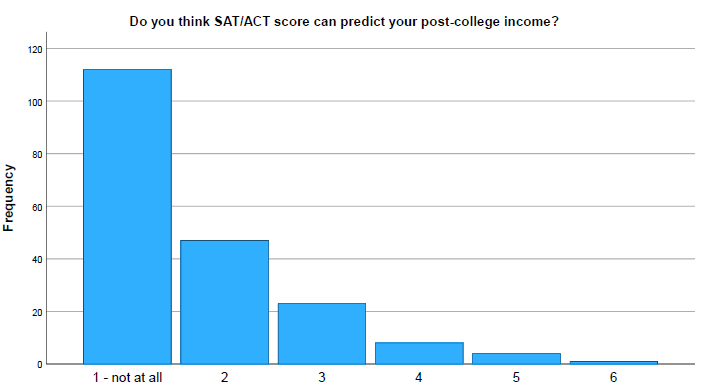}
    \caption{Student Response for Question 5}
    \label{fig:survey5_student}
\end{figure}

\renewcommand{\thefigure}{G.\arabic{figure}}
\setcounter{figure}{0}

\begin{figure}[htbp]
    \centering
    \includegraphics[width=\textwidth,keepaspectratio]{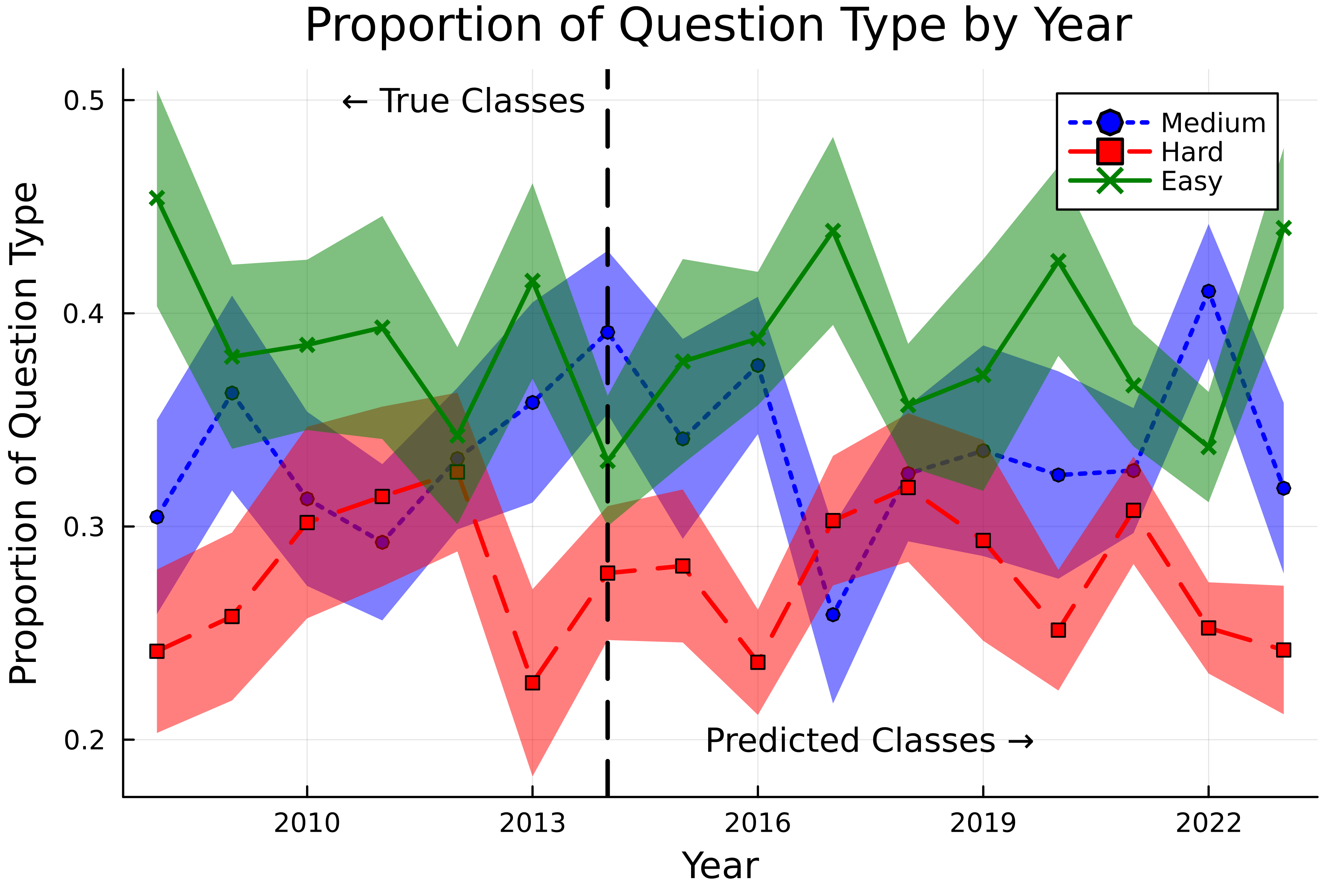}
    \caption{Difficulty Rating of SAT Questions in Bootstrapped Exams}
    \label{fig:difficulty_rating}
    \begin{minipage}{\textwidth}  
        \small
        \underline{Note}: The figure shows the proportion of easy, medium, and hard questions in the bootstrapped exams from each year. The difficulty of the questions is predicted using a Random Forest Classifier trained on the question embeddings and progress bar as features. The standard error is calculated using the proportion of predicted rating in the bootstrapped samples, it does not account for the uncertainty in the classifier for the unseen data between 2015 and 2023.
    \end{minipage}
\end{figure}

\begin{figure}[htbp]
    \centering
    \includegraphics[width=\textwidth,keepaspectratio]{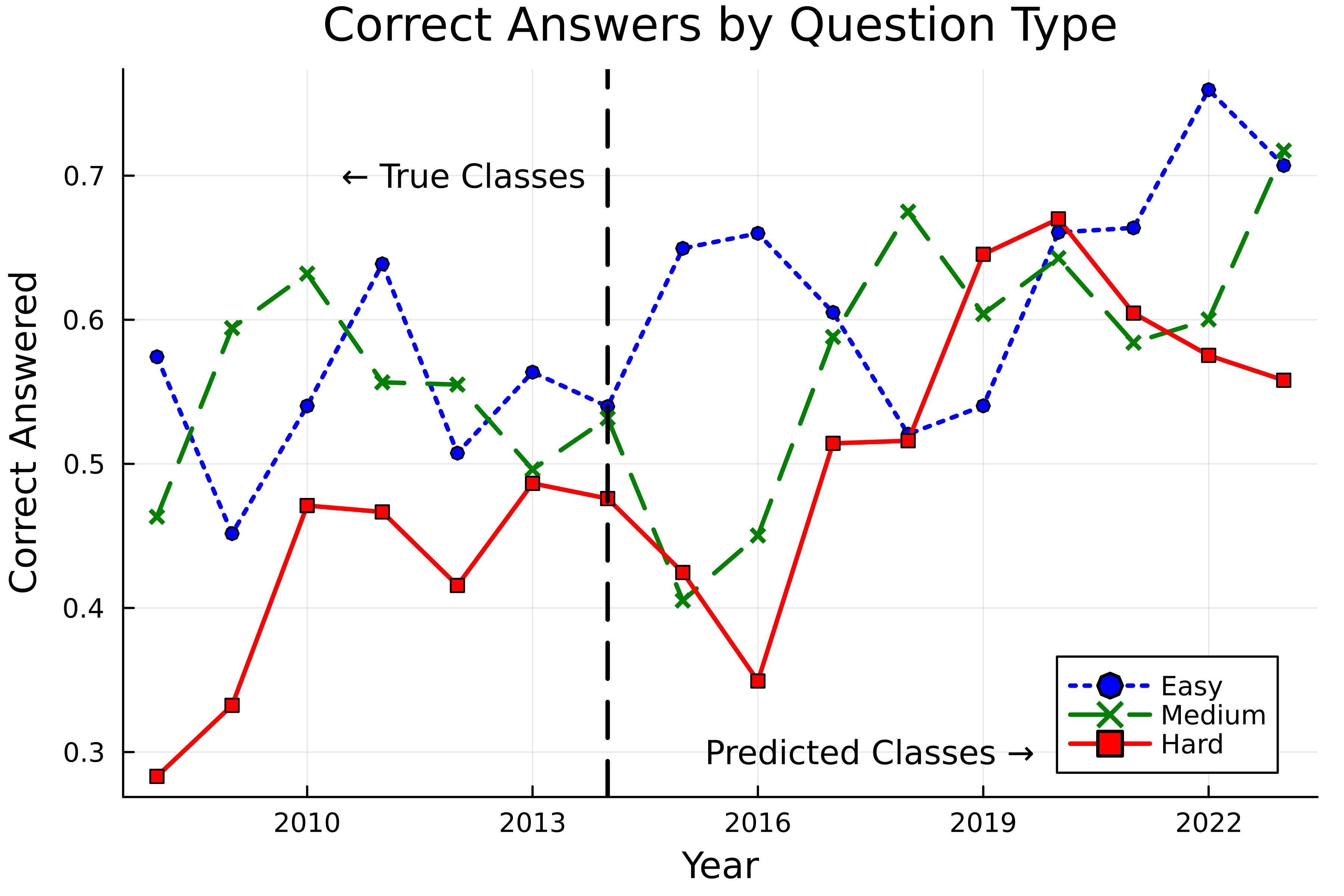}
    \caption{Difficulty Alignment of SAT Questions}
    \label{fig:difficulty_alignment}
\end{figure}

\pagebreak

\pagebreak

\end{document}